\documentclass[aps, prb, 10pt, twocolumn, superscriptaddress]{revtex4-2}
\usepackage[T1]{fontenc}
\usepackage[latin9]{inputenc}
\usepackage[dvipsnames]{xcolor}
\definecolor{shadecolor}{rgb}{1, 0, 0}
\usepackage{verbatim}
\usepackage{float}
\usepackage{framed}
\usepackage{amsmath}
\usepackage{amssymb}
\usepackage{graphicx}
\usepackage{physics}
\usepackage{cancel}
\usepackage{comment}
\usepackage{bm}

\usepackage{slashed}

\def\gradp{\boldsymbol{\nabla}_{\bf p}}
\def\gradr{\boldsymbol{\nabla}_{\bf r}}
\def\Om{{\boldsymbol\Omega}}

\newcommand{\del}{\partial}

\renewcommand{\epsilon}{\varepsilon}

\makeatletter
\PassOptionsToPackage{caption=false}{subfig}
\usepackage{hyperref}
\hypersetup{
breaklinks=true,
colorlinks=true,
citecolor=blue,
linkcolor=blue,
filecolor=blue,
urlcolor=blue
}
\IfFileExists{lmodern.sty}{\usepackage{lmodern}}{}
\makeatother

\usepackage{babel}

\begin{document}

\title{Nonlinear longitudinal current of band-geometric origin in wires of finite thickness}

\author{Robin Durand}
\author{Louis-Thomas Gendron}
\author{Th\'eo Nathaniel Dionne}
\author{Ion Garate}
\affiliation{D\'epartement de physique, Institut quantique and Regroupement Qu\'eb\'ecois sur les Mat\'eriaux de Pointe,  Universit\'{e} de Sherbrooke, Sherbrooke, Qu\'{e}bec J1K 2R1, Canada}

\date{\today}                                       

\begin{abstract}
The miniaturization of integrated circuits is facing an obstruction due to the escalating electrical resistivity of conventional copper interconnects. The underlying reason for this problem was unveiled by Fuchs and Sondheimer, who showed that thinner wires are more resistive because current- carrying electrons encounter the rough surfaces of the wire more frequently therein. Here, we present a generalization of the Fuchs-Sondheimer theory to Dirac and Weyl materials, which are candidates for next-generation interconnects. 
We predict a nonlinear longitudinal electric current originating from the combined action of the Berry curvature and non-specular surface-scattering.
\end{abstract}
\maketitle

{\em Introduction.--}
A strong research effort is currently underway to find new interconnect materials that will replace copper in integrated circuits \cite{gall2020,gall2021materials}. Topological materials, endowed with their robust and highly conducting surface states, have the potential \cite{zhang2019ultrahigh, breitkreiz2019large, gilbert2021topological, lanzillo2022size, lien2023unconventional} to overturn the unfavorable resistivity scaling predicted by Fuchs \cite{fuchs1938} and Sondheimer \cite{sondheimer1952} for conventional wires. 




In this work, we generalize the Fuchs-Sondheimer (FS) theory to materials with nonzero electronic Berry curvature. 
For a homogeneous wire with time-reversal symmetry, 
we predict an electric current of band-geometric origin along the wire axis, which is quadratic in the applied electric field. 
While nonlinear Hall effects produced by the Berry curvature \cite{sodemann2015quantum, ortix2021nonlinear} have been observed \cite{ma2019observation, kang2019nonlinear}, the nonlinear longitudinal current found below is fundamentally different. It emerges from the combined action of the Berry curvature and the non-specular scattering at the wire boundaries. 
Earlier studies \cite{ma2021topology, tsirkin2022separation, das2023intrinsic, wang2023intrinsic} of nonlinear currents associated to the quantum metric tensor have ignored wire boundaries, therefore overlooking the effect we predict. 


{\em Methodology.--}
We consider conducting materials with low-energy electronic valleys.
The electronic energy dispersion in the vicinity of valley $\chi$  is denoted as $\varepsilon_{\bf p}^\chi$, where ${\bf p}$ is the momentum measured from the valley center.
The system is finite in the $z$ direction, $z\in[-a/2,a/2]$, but infinite in the remaining directions. 
We assume that there is only one band (per valley) that intersects the Fermi energy, therefore (i) neglecting size quantization effects ($a$ must be much larger than the Fermi wavelength), and (ii) omitting the potential presence of topological surface states (we focus on the transport properties of bulk electrons).

The semiclassical transport under a uniform and constant electric field, ${\bf E}=E_x \hat{\bf x}$,
is described by the Boltzmann equation
\begin{equation}
\label{eq:boltzmann}
  \dot{\bf r}^\chi_{\bf p} \cdot \gradr f^\chi_{\bf p}({\bf r}) + \Dot{\bf p} \cdot \gradp f^\chi_{\bf p}({\bf r})= {\cal I}_{\bf p}^\chi,
\end{equation}
where $f^\chi_{\bf p}({\bf r})$ is the electronic occupation,
  \begin{align}
        \dot{\bf r}^\chi_{\bf p} &= {\bf v}^\chi_{\bf p}+  \Dot{\bf p}\times \Om^\chi_{\bf p}/\hbar\notag\\
        \Dot{\bf p} &= -e{\bf E}
    \end{align}
are the electronic group velocity and force, ${\bf v}^\chi_{\bf p}=\gradp \varepsilon^\chi_{\bf p}$, 
$\Om^\chi_{\bf p}$
is the Berry curvature in valley $\chi$  and 
\begin{equation}
\label{eq:coll}
{\cal I}_{\bf p}^\chi=-\frac{1}{\rho(\varepsilon_{\bf p}^\chi)\tau} \sum_{{\bf p}'} \delta(\varepsilon_{\bf p}-\varepsilon_{{\bf p}'}) (f^\chi_{\bf p}-f^\chi_{{\bf p}'})
\end{equation}
is the collision integral describing elastic scattering with bulk impurities in the lowest Born approximation.  Here, $\tau$ is the mean scattering time and $\rho(\epsilon)$ is the density of states at energy $\epsilon$. 
Treating the valleys independently is justified if the intravalley scattering time is much shorter than the intervalley scattering time, which we assume to be the case for the sake of simplicity. 
Likewise, we neglect skew-scattering and side-jump processes. 

We solve Eq.~(\ref{eq:boltzmann})  perturbatively in powers of $E_x$ up to second order, $f^\chi_{\bf p} \simeq f_{0,{\bf p}}^\chi + f^\chi_{1,{\bf p}} + f^\chi_{2,{\bf p}}$, where $f_{n,{\bf p}}^\chi \propto E_x^n$.
For $n\geq 1$, we apply the boundary condition of Fuchs and Sondheimer \cite{fuchs1938,sondheimer1952}: $f^\chi_{n,\bf p}(x,y,-a/2)=0$ if $v^\chi_{{\bf p}, z}>0$, and $f^\chi_{n,\bf p}(x,y,a/2)=0$ if $v_{{\bf p}, z}^\chi<0$.
Physically, the electron distribution on each valley is in equilibrium immediately after a collision with the wire boundary.
In the boundary conditions, $v^\chi_{{\bf p},z}$ appears instead of $\dot{z}^\chi_{\bf p}$; this is a consequence of solving  Eq.~(\ref{eq:boltzmann})  perturbatively in $E_x$.

Our quantity of interest is the longitudinal electric current density 
\begin{equation}
\label{eq:j}
    j_x({\bf r}) = \sum_\chi j_x^\chi({\bf r}) = -e  \sum_\chi \int \frac{d^d p}{(2\pi \hbar)^d} \, \dot{x}_{\bf p}^\chi\,  f^\chi_{\bf p}({\bf r}),
\end{equation}
where $d$ is the dimensionality of the system and $j_x^\chi$ is the valley-resolved current density along $x$.
Note that Eq.~(\ref{eq:j})  does not include the contribution from the orbital magnetic moment of electrons \cite{xiao2006}.
More precisely, we omit the the contribution to the current coming from the part of $f_{\bf p}^\chi$ that is isotropic in {\bf p}  \cite{sm_fsb}.
Yet, this omission does not change the main prediction of our paper.


Broken translational symmetry along $z$ implies $f^\chi_{\bf p}({\bf r})=f^\chi_{\bf p}(z)$ and thus
$j_x({\bf r}) =  j_x(z)$.
We write 
\begin{equation}
j_x(z)\simeq \sum_\chi \left[ j^\chi_{1,x}(z) + j^\chi _{2,x}(z)\right] = j_{1,x}(z) + j_{2,x}(z),
\end{equation}
where $j_{n,x}(z)\propto E_x^n$.
The effective linear ($\overline{\sigma}_1$) and nonlinear  ($\overline{\sigma}_2$) conductivities are obtained by averaging the current density over the wire thickness,
\begin{equation}
\label{eq:jav}
\overline{j_x} = \frac{1}{a} \int_{-a/2}^{a/2} dz j_x(z) \simeq \overline{\sigma}_1 E_x + \overline{\sigma}_2 E_x^2.
\end{equation}

{\em Solution of Eq.~(\ref{eq:boltzmann}).--}
At zeroth order in $E_x$, Eq.~(\ref{eq:boltzmann}) yields $f^\chi_{\bf p}=f^\chi_{0,{\bf p}}$ and $\partial_z f_{0,{\bf p}}(z) =0$.
Using this, the Boltzmann equation at first order in $E_x$ reads
\begin{equation}
\label{eq:b1}
v_{{\bf p}, z}^\chi \frac{\partial f^\chi_{1, {\bf p}}}{\partial z} - e E_x \frac{\partial f^\chi_{0, {\bf p}}}{\partial p_x} = {\cal I}_{1,{\bf p}}^\chi,
\end{equation}
where ${\cal I}_{1,{\bf p}}^\chi$ is given by Eq.~(\ref{eq:coll}) with $f\to f_1$.
The second term on the left hand side acts as a source term that forces $f^\chi_{1,{\bf p}}\neq 0$.
Assuming that $\varepsilon^\chi_{\bf p}$ is invariant under $p_x\to -p_x$,
the source term is odd in $p_x$.
Hence, we can decompose $f_1$ in a part that is even in $p_x$ and a part that is odd, only the latter being nonzero according to Eq.~(\ref{eq:b1}). 
Further imposing the FS boundary conditions, we reach 
\begin{equation}
\label{eq:f1}
    f_{1,{\bf p}}^\chi = \left\{\begin{array}{ll}  h_{1,{\bf p}}^\chi \left(1-e^{-(z+a/2)/v^\chi_{{\bf p}, z}\tau}\right) &\text{, } v^\chi_{{\bf p}, z}>0\\
    						h_{1,{\bf p}}^\chi \left(1-e^{-(z-a/2)/v^\chi_{{\bf p}, z}\tau}\right) &\text{,  } v^\chi_{{\bf p},z}<0
						\end{array}\right.,
\end{equation}
where
\begin{equation}
h_{1,{\bf p}}^\chi \equiv e E_x \tau\frac{\del f_{0,{\bf p}}^\chi}{\del p_x}.
\end{equation}
The electronic occupation becomes $z-$dependent at linear order in $E_x$. 
This position-dependence is significant only in the proximity to $z=a/2$ (for $v^\chi_{{\bf p},z}<0$) and $z=-a/2$ (for $v^\chi_{{\bf p},z}>0$). Far enough from the wire surfaces, Eq.~(\ref{eq:f1}) converges to the standard expression for infinite systems, $h_{1,{\bf p}}^\chi$, independent of $z$.
These results are identical to the ones obtained by Fuchs and Sondheimer \cite{fuchs1938,sondheimer1952}, i.e. unaffected by the Berry curvature of electrons. 

New results arise at second order in $E_x$, where the Boltzmann equation reads \cite{sm_fsb}
\begin{equation}	
\label{eq:boltz2}
  v_{{\bf p}, z}^\chi\frac{\del f_{2, {\bf p}}^\chi}{\del z}+\frac{e E_x\Omega^\chi_{{\bf p},y}}{\hbar}\frac{\del f_{1, {\bf p}}^\chi}{\del z}-eE_x\frac{\del f_{1, {\bf p}}^\chi}{\del p_x}= {\cal I}_{2,{\bf p}}^\chi.
    \end{equation}	
Here, the $y$-component of the Berry curvature appears in a source term for $f_2$.
Now, the two source terms  (involving $\partial_{z}f_{1}^\chi$ and $\partial_{p_x} f_{1}^\chi$, respectively) do not in general have the same parity under $p_x\to -p_x$. 
Therefore, if we decompose $f_2$ in a part that is even in $p_x$ and a part that is odd (once again under the assumption of $\epsilon_{\bf p}^\chi$ being an even function of $p_x$), both parts are nonzero according to Eq.~(\ref{eq:b1}). 
The collision term for the even part cannot be solved analytically without invoking approximations that violate conservation laws.
Fortunately, for the electric current in Eq.~(\ref{eq:j}), only the part of $f_2$ that is odd in $p_x$ matters 
\footnote{We have omitted any hypothetical internal electric field component in the $z$ direction. 
In systems displaying a nonlinear Hall effect, an internal nonzero $E_z$ would emerge, which would be quadratic in $E_x$. 
However, the source term produced by such a field in Eq.~(\ref{eq:boltz2}) would be even in $p_x$ and hence irrelevant for transport along $x$.}.
This part {\em can} be analytically obtained \cite{sm_fsb} without resorting to approximations:
 \begin{equation}
 \label{eq:f2_odd}
     f_{2,{\bf p}}^\chi \ni \left\{\begin{array}{ll}    h_{2, {\bf p}}^\chi (z+a/2) e^{-(z+a/2)/v_{{\bf p}, z}^\chi\tau} &\text{, } v_{{\bf p},z}^\chi>0\\
  h_{2, {\bf p}}^\chi  (z-a/2) e^{-(z-a/2)/v_{{\bf p}, z}^\chi \tau} &\text{, } v_{{\bf p}, z}^\chi<0,
    \end{array}\right.
    \end{equation}	
where
\begin{equation}
\label{eq:h2}	
h_{2, {\bf p}}^\chi \equiv  \frac{e^2 E_x^2 \Omega^\chi_{{\bf p},y}}{\hbar (v_{{\bf p}, z}^\chi)^2} \frac{\partial f^\chi_{0,{\bf p}}}{\partial p_x}. 
\end{equation}
Interestingly, Eq.~(\ref{eq:f2_odd}) is proportional to the $y-$component of the Berry curvature and emerges because $f_{1,{\bf p}}^\chi$ is $z-$dependent. Unlike $f_{1,{\bf p}}^\chi$, $f_{2,{\bf p}}^\chi$ is negligible far enough from the wire edges.
In our theory, the absence of an $O(E_x^2)$ current far from the edges is a consequence of the assumed invariance of $\epsilon_{\bf p}^\chi$ under $p_x\to -p_x$.  
When this assumption is relaxed, we find that there can be an additional $O(E_x^2)$ current density independent of Berry curvature for all $z$, provided that the crystal lacks time-reversal symmetry \footnote{This statement is in agreement with conventional theories (for a review, see e.g. Ref. [\onlinecite{gao2019semiclassical}]).}.
More so, should we incorporate skew-scattering processes in our theory, an $O(E_x^2)$ current density could likewise emerge for all $z$ even if $\epsilon_{\bf p}^\chi$ were invariant under $p_x\to -p_x$, and even in non-magnetic crystals \cite{isobe2020high}.

Equation~(\ref{eq:f2_odd}) foreshadows a nonlinear, surface-localized longitudinal current that originates from the joint presence of nontrivial band geometry and diffusive scattering at the wire's surfaces.
The precise form of that current is system-dependent. To make further progress, we next focus on Dirac and Weyl materials.


{\em Two dimensional Dirac metal.--}
We consider a crystal with time-reversal symmetry ${\cal T}$  and broken space inversion symmetry  ${\cal P}$ (e.g. a transition metal dichalcogenide monolayer) in the $xz$ plane, with $x\in(-\infty,\infty)$ and $z\in[-a/2,a/2]$.
We suppose that the Fermi energy $\epsilon_F$ intersects the conduction band, whose nondegenerate (due to broken ${\cal P}$)
dispersion at valley $\chi$ ($\chi=\pm 1$) is approximated as 
\begin{equation}
\label{eq:disp2D}
     \varepsilon^\chi_{\bf p} = \sqrt{c^2 p^2+m^2} + \alpha^\chi_z p_z.
\end{equation}
Here, $c$ is the Fermi velocity, ${\bf p}=(p_x, p_z)$ is the momentum measured from the band edge, $m$ is the Dirac mass and $\alpha^\chi_z$ is the tilt velocity along the $p_z$ direction. 
Since the two valleys are related by ${\cal T}$, we have $\alpha_z^\chi = \alpha_z \chi$ and a Berry curvature
\cite{xiao2010berry}
\begin{equation}
\label{eq:berry2D}
\Om_{\bf p}^\chi = \chi \frac{m \hbar^2 c^2}{2 \left(c^2 p^2 + m^2\right)^{3/2}} \hat{\bf y}.
\end{equation}
In addition, the model and the FS boundary conditions display a mirror symmetry ${\cal M}_z$ about a plane perpendicular to $z$.

We compute $j_{2, x}^\chi(z)$, the contribution from valley $\chi$ to the nonlinear part of the longitudinal current, by combining Eqs.~(\ref{eq:berry2D}), (\ref{eq:f2_odd}) and (\ref{eq:j}).  
To zeroth order in $\alpha_z$ and at low temperature, we obtain \cite{sm_fsb}
\begin{align}
\label{eq:j2_2D}
      &  j_{2, x}^\chi(z) \simeq \chi E_x^2 \frac{e^3}{8\pi^2\hbar} \frac{m}{\epsilon_F^2}  \int_0^{\pi} \frac{d \theta}{\tan^2\theta} \notag\\
 &\times\left[\left(z+\frac{a}{2}\right)e^{-\frac{z+a/2}{l\sin{\theta}}}-\left(-z+\frac{a}{2} \right) e^{-\frac{-z+a/2}{l\sin{\theta}}} \right],
\end{align}
where $l=c\tau (1-m^2/\epsilon_F^2)^{1/2}$ is an effective bulk mean free path due to impurity scattering.

Four properties of Eq.~(\ref{eq:j2_2D}) are worth mentioning. 
First, $j_{2, x}^+ (z) = j_{2, x}^- (-z)$, which is a consequence of the mirror symmetry ${\cal M}_z$. The same property is satisfied at first order in $E_x$, i.e. $j_{1, x}^+ (z) = j_{1, x}^- (-z)$.
As a consequence, $j_x(z) = \sum_\chi j_x^\chi(z) = j_x(-z)$ at all orders in $E_x$, which is expected because an electric field along $x$ preserves ${\cal M}_z$. 

Second, $j_{2, x}^\chi (z)=-j_{2, x}^\chi (-z)$.
This property can be understood mathematically from the fact that, in the absence of a tilt, the electronic dispersion (\ref{eq:disp2D}) and the Berry curvature (\ref{eq:berry2D}) of each valley are invariant under $(x, z; p_x, p_z)\to (-x, -z; -p_x, -p_z)$.
Under this operation, $j_{2, x}^\chi (z) = \sigma_2 E_x^2 \to -j_{2, x}^\chi(-z) = \sigma_2' E_x^2$.
From Neumann's principle \cite{powell2010a}, $\sigma_2=\sigma_2'$ and therefore $j_{2, x}^\chi (z) =-j_{2, x}^\chi (-z)$. 
In contrast, $j_{1, x}^\chi(z)=\sigma_1 E_x \to -j_{1, x}^\chi(-z)=\sigma_1' (-E_x)$. 
Here, the symmetry condition $\sigma_1=\sigma_1'$ leads to  $j_{1, x}^\chi(z)=j_{1, x}^\chi(-z)$.

Third, $j_{2,x}^\chi(z)$ is localized near the edges of the wire.
Such localization is directly associated to the second property, $j_{2, x}^\chi (z)=-j_{2, x}^\chi (-z)$.
Indeed, far from the wire edges, the current density is expected to be spatially uniform and thus independent of $z$; the relation $j_{2, x}^\chi (z)=-j_{2, x}^\chi (-z)$ then results in  $j_{2, x}^\chi=-j_{2, x}^\chi = 0$ far from the edges. 
We emphasize that this localization of $j_{2, x}(z)$ near the edges takes place for bulk electrons, as our theory does not contain any surface-localized electronic bands. In addition, the localization length for $j_{2,x}^\chi(z)$ is of the order of the bulk mean free path, unlike what one would expect from surface-states.  Such spatial distribution of $j_{2,x}^\chi(z)$ is contrary to that of $j_{1,x}^\chi(z)$: at first order in $E_x$, the current is distributed along the entire bulk of the wire (even far from the edges, as allowed by relation $j_{1,x}^\chi(z)=j_{1,x}^\chi(-z)$)  and partially depleted at the surfaces of the wire due to diffusive scattering therein.

Fourth, $j_{2, x}^\chi(z)$ is equal in magnitude and opposite in sign for the two valleys. 
This is an immediate joint consequence of the first two properties.
Therefore, $j_{2, x}(z) = \sum_\chi j_{2, x}^\chi(z)=0$ for all $z$.
This is again different to the situation at first order in $E_x$, where $j_{1, x}^\chi (z)=j_{1, x}^\chi (-z)$ and hence the two valleys contribute with equal magnitude and sign when $\alpha_z=0$.

\begin{figure}[t]
    \includegraphics[width=\hsize]{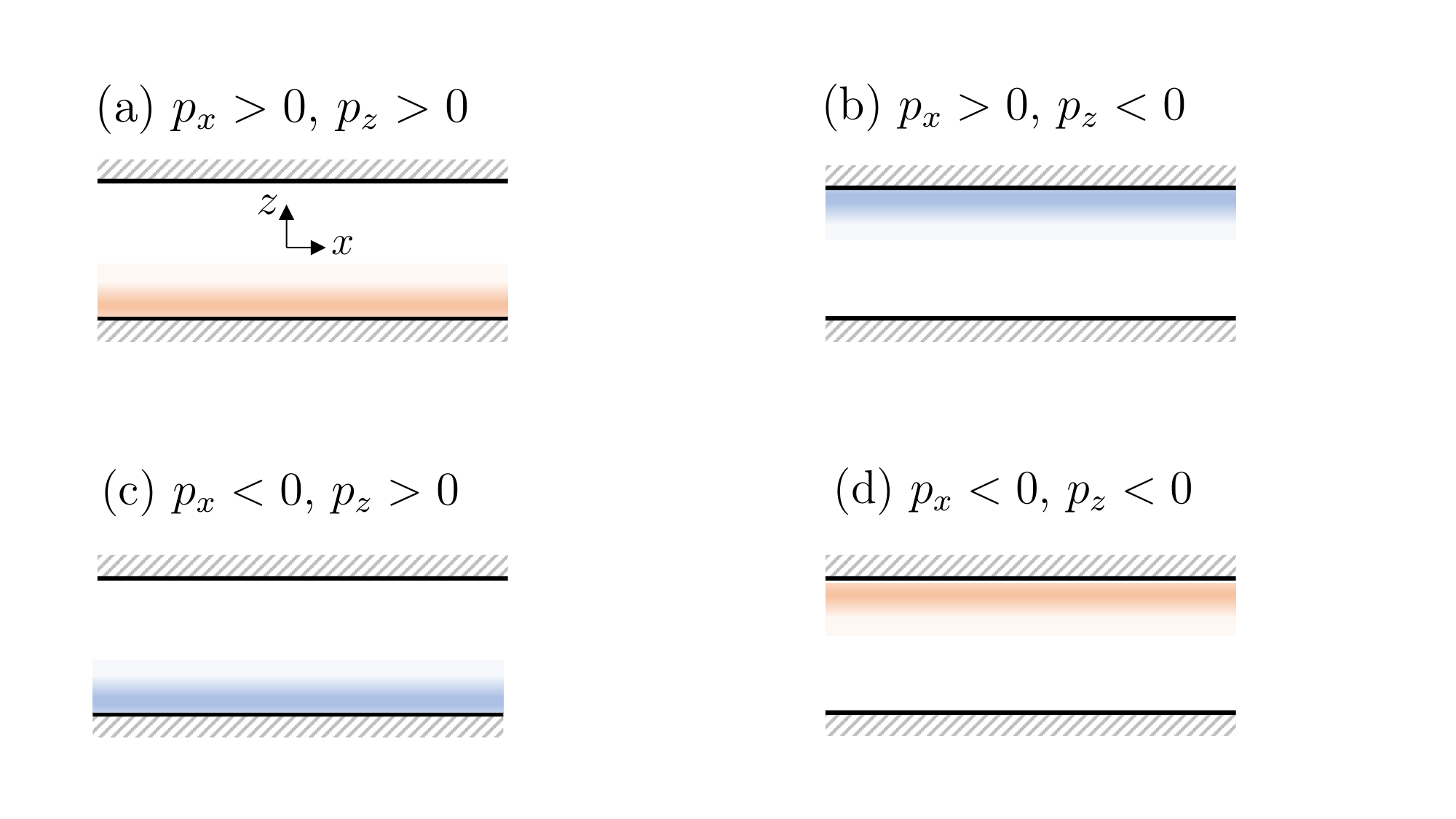}
    \caption{Schematic contribution of the anomalous velocity to the electronic distribution function at a single valley of a 2D metal, as a function of the electronic momenta $(p_x, p_z)$. Orange: accumulation; blue: depletion; white: negligible contribution. Black solid lines denote the surfaces of the wire; grey dashed lines denote regions outside the wire. Regardless of  $p_z$, there is an excess of right-moving electrons ($p_x>0$) on one side of the wire and an excess of left-moving electrons ($p_x<0$) on the opposite side of the wire.}   
    \label{fig:cartoon}
\end{figure}

A physical picture for the second and third properties is sketched in Fig.~\ref{fig:cartoon}. 
The Berry curvature leads to an anomalous velocity $\propto E_x \Omega^\chi_{{\bf p}, y}$ along the $z$ direction
for electrons of momentum ${\bf p}$ at valley $\chi$. 
Consequently, the term $E_x \Omega^\chi_{{\bf p}, y} \partial_z f_{1,{\bf p}}^\chi$ in Eq.~(\ref{eq:boltz2})  describes the leading order contribution of the anomalous velocity to the divergence of the electronic current density  ($f^\chi_{1,{\bf p}}$ appears to leading order, because $\partial_z f^\chi_{0,{\bf p}}=0$). Here, the word ``contribution'' is to be emphasized,
as the total current density along $z$ is zero (i.e., the integral of $E_x \Omega^\chi_{{\bf p}, y}  f^\chi_{1,{\bf p}}$ over ${\bf p}$ vanishes). But, if we focus on electrons with momentum ${\bf p}$, their flow has a nonzero divergence in the vicinity of the wire edges (the flow is divergenceless far from the edges because $\partial_z f^\chi_1\simeq 0$ therein according to Eq.~(\ref{eq:f1})).
This implies an accumulation of electrons of a given ${\bf p}$ near the wire edges.

To determine the nature of such accumulation, we notice that (i) for fixed $p_z$, $f^\chi_{1,{\bf p}}$ has opposite sign for $p_x>0$ and $p_x<0$ (this is expected as there is a net electric current along $x$ at first order in $E_x$); (ii) for $p_z>0$, $\partial_z f^\chi_{1,{\bf p}}$ is nonzero only near $z=-a/2$, while for $p_z<0$,  $\partial_z f^\chi_{1,{\bf p}}$ is nonzero only near $z=a/2$ (cf. Eq.~(\ref{eq:f1})); (iii) for fixed $p_x$, $\partial_z f^\chi_{1,{\bf p}}$ switches sign under $p_z\to -p_z$ (this is again a consequence of the FS boundary conditions).
It follows from (i), (ii) and (iii) that electrons with $p_x>0$ and $p_x<0$ accumulate on opposite edges of the wire (cf.  Fig.~\ref{fig:cartoon}). The accumulation takes place at order $E_x^2$ (one factor of $E_x$ comes from the anomalous velocity, and the other from the spatial gradient of the electronic distribution function caused by diffusive boundary scattering).
This is equivalent to a current distribution, which is quadratic in $E_x$, localized at the surfaces of the wires and antisymmetric with respect to the center of the wire. 
Hence, the second and third properties mentioned above are realized.

\begin{figure}[t]
    \includegraphics[width=\hsize]{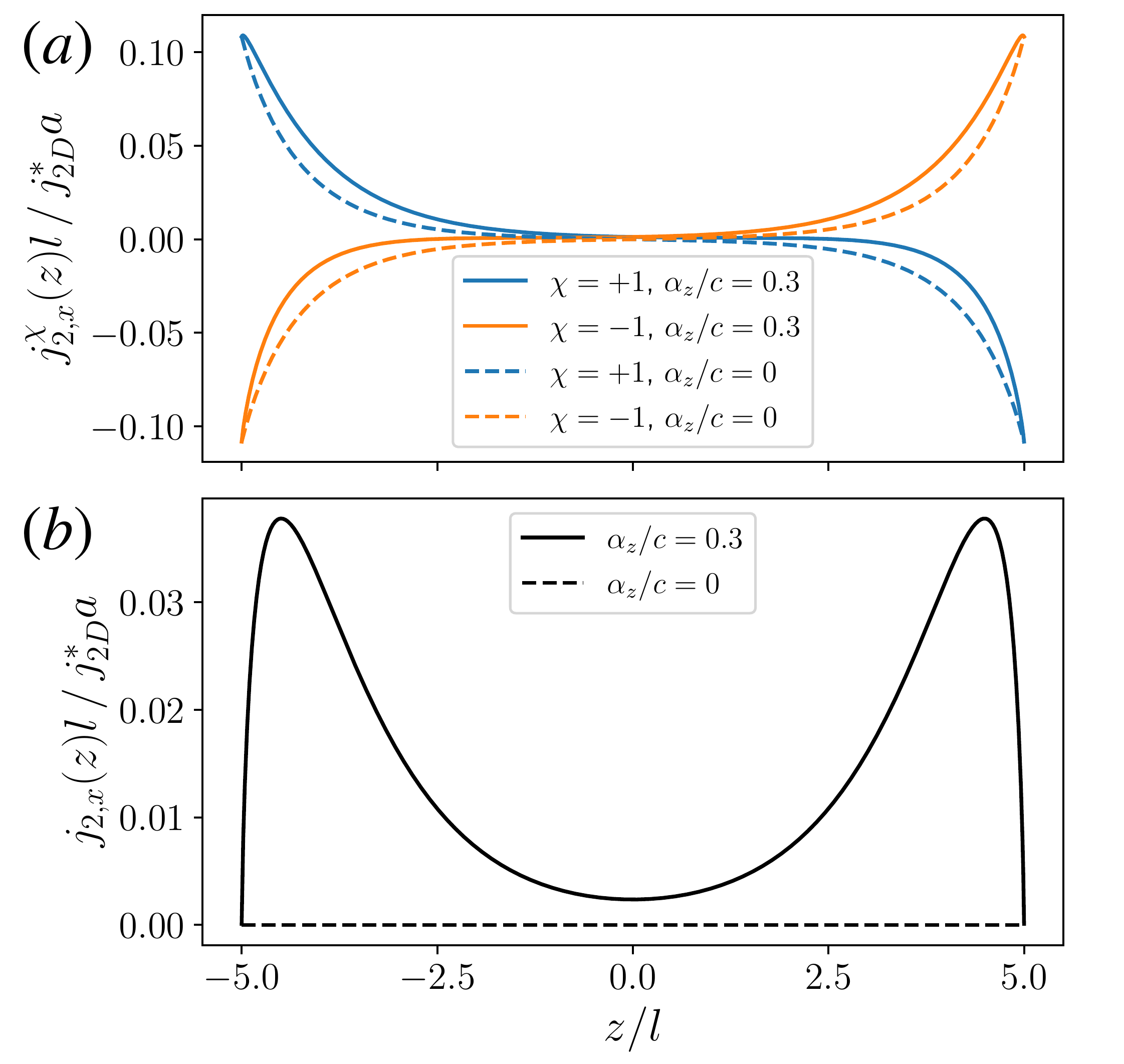}
    \caption{Nonlinear longitudinal current density caused by the Berry curvature in a time-reversal-symmetric 2D Dirac metal with two electronic valleys ($\chi=\pm 1$). We take $a = 10l$ and $m=0.5\varepsilon_F$. A tilt-independent normalization current $j^*_{2D} = j_{2D}c/\alpha_z$ is introduced, where $j_{2D}$ is defined in the main text.  (a) The valley-resolved current density $j_{2, x}^\chi(z)$  is localized near the wire edges, with a localization length of the order of $l$. In the absence of a tilt of the energy dispersion in the $z$ direction ($\alpha_z=0$), $j_{2, x}^\chi(z)$ is antisymmetric with respect to the center of the wire, and takes opposite values for the two valleys. A tilt $\alpha_z\neq 0$ skews the current density profile and prevents the cancellation between the two valleys. (b) The valley-summed nonlinear longitudinal current density $j_{2, x}(z)$ vanishes at every $z$ when $\alpha_z=0$. When $\alpha_z\neq 0$, $j_{2, x}(z)$ is nonzero and localized at the wire edges.\label{fig:j2}}
\end{figure}

The presence of a tilt of the electronic structure along $p_z$ changes some of the preceding properties qualitatively. 
Crucially, $\alpha_z\neq 0$ breaks the antisymmetry of $j^\chi_{2,x}(z)$ 
about the center of the wire (see Fig.~\ref{fig:j2}). This is because  $(x, z; p_x, p_z)\to (-x, -z; -p_x, -p_z)$ is no longer a symmetry when $\alpha_z\neq 0$.
As a result, the two valleys no longer make mutually cancelling contributions to the nonlinear longitudinal current and $j_{2, x}(z)$ becomes nonzero.
The relations $j_{n, x}^+(z)=j_{n, x}^-(-z)$ and $j_{n,x}(z)=j_{n,x}(-z)$ continue to hold for $n=1,2$ because the tilt along $p_z$ preserves the mirror symmetry ${\cal M}_z$.
Also, $j_{n, x}^\chi(z)$ and by extension $j_{2, x}(z)$ continue to be localized near the wire surfaces.
In Ref.~\cite{sm_fsb},
we show an expression for $j_{2,x}(z)$ that is valid up to first order in $\alpha_z/c$. This term scales as the product of $\alpha_z$ and the Berry curvature.

Combining the expression for $j_{2,x}(z)$ in \cite{sm_fsb}
with Eq.~(\ref{eq:jav}), we obtain the nonlinear part of the current density averaged over the thickness of the wire at low temperature:
\begin{equation}
\label{eq:mainresult1}
       \overline{j_{2,x}} =\overline{\sigma}_2 E_x^2 \simeq j_{\rm 2D} \left(1- I\right),
\end{equation}
where 
\begin{equation}
I=\frac{3}{8}\int_{0}^{\pi} d\theta \frac{\cos^2\theta}{\sin\theta}\left(\frac{a}{l}+2 \sin\theta\right)^2\exp\left(-\frac{a}{l\sin{\theta}}\right)
\end{equation}
is a dimensionless integral and
\begin{equation}
j_{\rm 2D}\equiv E_x^2 \frac{16}{3} \frac{ e^3}{\pi^2\hbar^3 }\frac{\epsilon_F\Omega_y l^2}{ c^2 a} \sqrt{1-\frac{m^2}{\epsilon_F^2}}\frac{\alpha_z}{c}
\end{equation}
is proportional to the Berry curvature evaluated at the Fermi energy at valley $\chi=+1$, i.e. $\Omega_y = m\hbar^2c^2/2\epsilon_F^3$.
This current reverses sign  when time-reversal is applied to the system via $\alpha_z\to -\alpha_z$, $m\to-m$, $\tau\to -\tau$, $c\to -c$, with the other quantities remaining invariant.  
According to conventional knowledge \cite{gao2019semiclassical}, there should not exist a current proportional to $\tau^2$ in a crystal with ${\cal T}$, at least if skew-scattering is ignored \cite{isobe2020high}. Yet, such statement is no longer applicable when the electronic occupation varies in space, like in our case.

Analytical expressions of  Eq.~(\ref{eq:mainresult1}) are accessible in limiting regimes:
 \begin{equation}
 \label{eq:lim2D}
   \overline{j_{2,x}}\simeq \left\{\begin{array}{ll} j_{\rm 2D}   &\text{, } a\gg l\\
   9 a^2/(4 l^2) \ln (l/a) j_{2 D}    &\text{, } a\ll l.
    \end{array}\right.
    \end{equation}	
The fact that $\overline{\sigma}_2\propto 1/a$ when $a\gg l$ reflects the spatial localization of  $j_{2,x}(z)$  near the wire edges, with a localization length that is independent of $a$.
Thus, $\overline{\sigma}_2\to 0$ when $a\to \infty$.
Likewise, $\overline{\sigma}_2\to 0$ in the  $a\to 0$ limit, when  the diffusive boundary scattering becomes pervasive.
Given that $\overline{j_{2,x}}$ vanishes when $a\to 0$ and $a\to \infty$, it must have a maximum somewhere in between. 
The maximum of $\overline{\sigma}_2$, of value $\sim j_{2D}/E_x^2$,  takes place at $a\sim l$.
For $a\sim l$, $m\simeq 100 {\rm meV}$, $c\simeq 10^5 {\rm m/s}$ and $\alpha_z/c\simeq 0.3$, we find $\overline{\sigma}_2/\overline{\sigma}_1\simeq 10^{-3}\mu \text{m.volt}^{-1}$
\footnote{To leading order in $\alpha_z$, $\overline{\sigma}_2/\overline{\sigma}_1 \simeq e h \alpha_z/(6\pi m^2)$ for $a\simeq l$.}
. Such ratio is similar to the one reported in the nonlinear Hall conductivity experiments (with the important difference that $\sigma_2$ is a longitudinal conductivity in our case). 


{\em Three dimensional Weyl metal.--}
Next, we consider a Weyl semimetal (WSM) \cite{armitage2018weyl}  thin film in the $xy$ plane, with $x,y\in(-\infty,\infty)$,  and $z\in[-a/2,a/2]$.
We start with two valleys of opposite chirality $\chi=\pm 1$, related by an improper crystal symmetry.
We suppose that the Fermi energy $\epsilon_F$ intersects the conduction band.
The
electronic dispersion at valley $\chi$ is approximated as 
\begin{equation}
\label{dispersion}
     \varepsilon^\chi_{\bf p} = c p+ \alpha_y^\chi p_y +\alpha_z^\chi p_z,
\end{equation}
where $c$ is the Fermi velocity, ${\bf p}=(p_x, p_y, p_z)$ is the momentum measured from the Weyl point, $\alpha^\chi_y$ and $\alpha^\chi_z$ are tilt velocities at valley $\chi$ in the $y$ and $z$ directions, respectively. 
The Berry curvature for this model reads
\begin{equation}
\label{eq:berry_weyl}
\Om_{\bf p}^\chi = \chi \hbar^2 {\bf p}/(2 p^3).
\end{equation}

Combining Eqs.~(\ref{eq:berry_weyl}),  (\ref{eq:f2_odd}) and (\ref{eq:j}), we compute $j_{2, x}^\chi(z)$ \cite{sm_fsb}.
Like in two dimensions, $j_{2, x}^\chi(z)$  is exponentially localized at the boundaries of the wire.
In addition, if $\alpha_z^\chi=0$, we find $j_{2, x}^\chi(z)=-j_{2, x}^\chi(-z)$, again like in the 2D Dirac metal.
We notice, however, that $\alpha_y^{\chi}\neq 0$ is necessary for $j_{2, x}^\chi(z)\neq 0$, because $\Om_{{\bf p},y}^\chi$ is an odd function of $p_y$ in a Weyl semimetal.


Averaging the current density over the wire thickness, we get $ \overline{j_{2,x}}=0$ when either $\alpha_y^\chi$ or $\alpha_z^\chi$ are zero \cite{sm_fsb}.
Thus, it is required that the Weyl cones be tilted in both directions transverse to the transport direction.
This is analogous to what we found for a 2D Dirac metal.
The two valleys make perfectly cancelling contributions to $\overline{j_{2, x}}$ if $\alpha_y^\chi \alpha^\chi_z$ is independent of $\chi$; such is the case when the two valleys are related to one another by a mirror plane perpendicular to the $x-$axis ($\alpha^+_z=-\alpha^-_z$ and $\alpha^+_y = -\alpha^-_y$). 
If the two valleys are instead related by a mirror plane perpendicular to the $z$ axis, then we have $\alpha^+_z=-\alpha^-_z$ and $\alpha^+_y = \alpha^-_y$. In such case, $j_{2, x}^+(z) = j_{2, x}^-(-z)$ and $\overline{j_{2,x}}\neq 0$.

In thick and thin films, we get 
 \begin{equation}
 \label{eq:lim3D}
   \overline{j_{2,x}}\simeq \left\{\begin{array}{ll} 
    j_{\rm 3D}   &\text{, } a\gg l\\
    3 a^2/(2 l^2) \ln (l/a) j_{3 D} 
    &\text{, } a\ll l
    \end{array}\right.
    \end{equation}	
   to leading order in $|\alpha_y^\chi|/c$ and $|\alpha^\chi_z|/c$, with
    \begin{equation}
    \label{eq:j3D}
    j_{\rm 3D}\equiv E_x^2 \frac{e^3l^2}{12 a  h^2 c} \sum_\chi \chi \frac{\alpha_y^\chi \alpha_z^\chi}{c^2}.
    \end{equation}   
Thus, $\overline{j_{2,x}}$ vanishes when $a\to\infty$ and $a\to 0$, its maximum of value $j_{3D}/2$ occurring when $a\sim 0.5l$.
For $\alpha_z^\chi/c \simeq 0.3 \chi$, $\alpha_y^\chi/c\simeq 0.3$ and $a\simeq 0.5 l$ , we get $\overline{\sigma}_2 \simeq  10^{-6} l[{\rm nm}] {\rm ohm}^{-1}{\rm volt}^{-1}$.
In comparison, a nonlinear longitudinal conductivity of  $10^{-3} {\rm ohm}^{-1}{\rm volt}^{-1}$ has been recently measured in chiral tellurium  \cite{suarez2024odd}.

The preceding analysis can be extended to WSM with multiple pairs of valleys. 
For example, in a WSM with time-reversal symmetry, 
valleys related by ${\cal T}$ have opposite tilts but the same chirality.
Based on Eq.~(\ref{eq:j3D}), those valleys make equal contributions to $\overline{j_{2,x}}$, thereby enhancing it.


{\em Conclusion.--}
We have unveiled a mechanism for 
a nonlinear longitudinal current in non-magnetic wires of finite thickness, resulting from 
(i) non-specular scattering at wire surfaces; (ii) a Berry curvature of electrons in the direction orthogonal to both the transport and the wire surface; 
(iii) an energy dispersion that is asymmetric at each electronic valley along the directions orthogonal to transport.
Conditions (i) and (ii)  ensure a surface-localized current distribution, {\em quadratic} in the applied electric field and parallel to the field.
Condition (iii) enables a nonzero average of the nonlinear current density over the thickness of the wire. Skew-scattering and side-jump are not required.

We have illustrated our theory for Dirac and Weyl materials, where the effect can be sizeable.
An experimental challenge is to distinguish the surface current predicted here from more conventional \cite{isobe2020high} but symmetry-allowed nonlinear bulk currents.
Possible strategies include thickness- and electron-density-dependent conductance measurements or the use of local current probes \cite{nowack2013imaging} (see End Matter for an extended discussion).

Our study was limited to a wire that is spatially homogeneous in equilibrium.
In real interconnects, the composition of the wire along its cross section is inhomogeneous. In such case, an inspection of Eq.~(\ref{eq:boltzmann}) shows that the Berry curvature  contributes to the longitudinal electric current at {\em first order} in the applied electric field. Although further work is needed to quantify this scenario,  there is a prospect to engineer smooth electron concentration gradients in interconnects in order to enhance their conductivity by virtue of the Berry curvature.
Another future direction of research is to generalize our theory to the situation where the bulk states coexist with surface states, and intervalley scattering is not negligible.

{\em Acknowlegdements.-}
This work has been financially supported by the Canada First Research Excellence Fund, the Natural Sciences and Engineering Research Council of Canada (Grant No. RGPIN- 2018-05385).
We are grateful to C.-T. Chen, E. Lantagne-Hurtubise and A.-M. Tremblay for informative discussions.

\twocolumngrid
\bibliography{biblio}{}
\bibliographystyle{apsrev4-1}

\onecolumngrid

\section*{End Matter}
\subsection*{Discussion: experimental detection}

\twocolumngrid
The main prediction of this work is a longitudinal, nonlinear, surface-localized current that is produced by the combined action of the bulk Berry curvature and diffusive boundary scattering. 
As  we discuss next, there are challenges to unequivocally detect such current, but there may be viable paths ahead.

Overall, nonlinear currents can be experimentally distinguished from the dominant linear current, as done e.g. in Ref. \cite{suarez2024odd}.
Likewise, the spatial distribution of surface-localized currents can be mapped through SQUID measurements (see e.g. Ref. \cite{nowack2013imaging}). 

A more fundamental challenge for the detection of our proposed current is that it coexists with {\em bulk} currents (i.e. currents that are distributed over the entire volume of the sample) predicted in Ref. \cite{isobe2020high}. 
The latter originate from the mechanism of skew-scattering, which appears in the collision term of the Boltzmann equation beyond our lowest Born approximation.
At first thought, one could differentiate our surface-localized current from the bulk currents by measuring the scaling of the conductance with the wire thickness. If the wire is thick compared to the bulk mean free path, the contribution from the bulk currents of Ref. \cite{isobe2020high} to the measured conductance will scale linearly with the wire thickness. In contrast, our predicted contribution to conductance will be independent of the thickness. 

There is a caveat, however. 
While the theory of Ref. \cite{isobe2020high} was done for an infinite sample, it is likely that in finite wires it will be affected by diffusive boundary scattering. As a result, the contribution from skew-scattering to conductance may not be a perfectly linear function of the wire thickness.
In this case,  further analysis is required to distinguish the Berry curvature contribution to the current from the boundary-induced changes of the skew-scattering contribution to the current. 
To make matters more difficult, both currents have the same dependence on impurity concentration and on the scattering time.
Fortunately, the dependence of the current on the carrier density is rather different depending on the mechanism at hand (Berry curvature vs skew-scattering).
This is evident from comparing our expressions with those of Ref. \cite{isobe2020high}. Still, varying the electron density in a controllable way is not simple, especially in three dimensional  materials. 

An alternative approach consists of adopting a transport configuration in which bulk nonlinear longitudinal currents are forbidden by symmetry.
Specifically, let us discuss the case of a transition metal dichalcogenide monolayer in the $xy$ plane (the crystal orientations corresponding to $x$ and $y$ axes are illustrated in Ref. \cite{isobe2020high}; note that this nomenclature differs from the one we adopted in the main text, where the $y$ direction was out-of-plane.)

In this system, a bulk nonlinear longitudinal current is allowed by symmetry when the electric field points along $y$, but not when it points along $x$.
In the latter case, the energy dispersion at each valley is symmetric in the transverse direction ($\alpha_z^\chi=0$ in Eq. (\ref{eq:disp2D})); this is the microscopic origin for the vanishing of the nonlinear longitudinal current along $x$.


Importantly, our calculations (Eqs. (\ref{eq:f2_odd}), (\ref{eq:h2}) and Fig. \ref{fig:j2}a) show the existence of  nonlinear currents produced by the Berry curvature at each electronic valley, even when the electric field points in the $x$-direction. These valley-resolved currents are spatially localized within a mean free path from the wire boundaries. But, at each boundary, the two valleys make mutually cancelling contributions to the nonlinear current along $x$.

Now, we may avoid such cancellation between valleys by breaking time-reversal symmetry at (and only at) the wire boundaries, e.g. with magnetic interfaces. 
In the theory of Fuchs and Sondheimer, this will result in a boundary condition that is no longer the same for the two valleys. 
Consequently, the cancellation between the valleys will be prevented and net currents produced by the Berry curvature will flow along $x$ close the boundaries. In contrast, skew-scattering contribution to nonlinear current along $x$ is expected to vanish separately at each valley, due to the very same symmetry that prevents bulk nonlinear currents ($\chi_{xxx}^{(1)}=0$ in Ref. \cite{isobe2020high}).
Moreover, the effect of boundary scattering  being unimportant far from the boundary, the bulk currents will remain null therein.
This way, one can experimentally confirm our predicted current by measuring a nonlinear conductance of the wire that is independent of the wire thickness. 

A final remark is in order here. Since the energy dispersion along the transport direction $x$ is asymmetric in each valley, the time-reversal breaking boundary condition can generate an extra nonlinear current along $x$ in the vicinity of the boundary, which is unrelated to both skew-scattering and to the Berry curvature (see e.g. Eq. (S46) in Ref. \cite{isobe2020high}).
It is however feasible to distinguish such extra current from the one that emerges from the Berry curvature. The key point that allows to do so is that the Berry curvature contribution to the current at a given valley changes sign between the top and bottom boundaries of the wire (see Fig. \ref{fig:j2}a), whereas the non-Berry contribution does not. 
Thus, by breaking time-reversal symmetry oppositely (with same magnitude but opposite sign) on the top and bottom boundaries, one can single out the geometric (Berry) contribution to the nonlinear longitudinal current.

\newpage

\onecolumngrid
\section*{Supplementary Material: Nonlinear longitudinal current of band-geometric origin in wires of finite thickness}

\setcounter{equation}{0}
\setcounter{figure}{0}
\setcounter{table}{0}
\setcounter{page}{1}
\makeatletter
\renewcommand{\theequation}{S\arabic{equation}}
\renewcommand{\thefigure}{S\arabic{figure}}
\renewcommand{\thetable}{S\arabic{table}}


{\bf Outline}

Sec. A: Details about the solution of the Boltzmann equation

Sec. B: Derivation of the nonlinear current density in a two-dimensional Dirac metal

Sec. C: Derivation of the nonlinear current density in a three-dimensional Weyl semimetal

Sec. D: Valley polarization

Sec. E: Edge charge accumulation

Sec. F: Comparison with earlier literature on band-geometric currents in spatially inhomogeneous systems

\appendix

\section{Sec. A: Details about the solution of the Boltzmann equation}
The Boltzmann equation we wish to solve has the form
\begin{equation}
	\label{eq:boltz_app}
	\dot{z}_{\bf p} \pdv{f_{\bf p}^\chi}{z}-eE_x\pdv{f_{\bf p}^\chi}{p_x} =  -\frac{1}{\tau}\left( f_{\bf p}^\chi - \frac{1}{\rho(\varepsilon_{\bf p}^\chi)} \sum_{{\bf p}'}\,f_{{\bf p}'}^\chi\delta(\varepsilon_{\bf p}-\varepsilon_{\bf p'})\right),
\end{equation}
where $\tau$ is the elastic scattering time (assumed to be a constant), $\rho(\epsilon)$ is the electronic density of states at energy $\epsilon$ and the sum over ${\bf p}'$ is confined to the valley $\chi$.
Without inelastic scattering, steady state is reached only in the presence of a temperature gradient, needed to evacuate the Joule heating produced by the current \cite{tremblay1979joule1, tremblay1982joule2}. This temperature gradient is quadratic in $E_x$ and influences the part of the electron occupation that is isotropic in momentum. It will therefore not affect our calculation of the current and we will hence ignore it.

We solve Eq.~(\ref{eq:boltz_app})  perturbatively in powers of $E_x$ up to second order, $f^\chi_{\bf p} \simeq f_{0,{\bf p}}^\chi + f^\chi_{1,{\bf p}} + f^\chi_{2,{\bf p}}$, where $f_{n,{\bf p}}^\chi \propto E_x^n$.  We wish to avoid using the relaxation time approximation, which leads to a violation of conservation laws. 
We are able to do so, at the expense of introducing some simplifying assumptions.

To first order in $E_x$, Eq.~(\ref{eq:boltz_app}) reads
\begin{equation}
	\label{eq:boltz_app1}
	v_z^\chi \pdv{f_{1,{\bf p}}^\chi}{z}-e E_x\pdv{f_{0,{\bf p}}^\chi}{p_x} =  -\frac{1}{\tau}\left( f_{1,{\bf p}}^\chi - \frac{1}{\rho(\varepsilon_{\bf p}^\chi)} \sum_{{\bf p}'}\,f_{1,{\bf p}'}^\chi\delta(\varepsilon_{\bf p}-\varepsilon_{\bf p'})\right).
\end{equation}
To make further progress, we assume that $\epsilon_{\bf p}^\chi$ is invariant under $p_x\to -p_x$.
Then, the second term in left hand side in Eq.~(\ref{eq:boltz_app1}), which acts as a source term for $f_{1,{\bf p}}^\chi$, is an odd function of $p_x$. 
As a result, if we make an ansatz of separating $f_{1,{\bf p}}^\chi$ into a part that is even in $p_x$ and a part that is odd in $p_x$, then the even part can be taken to be equal to zero. 
For the odd part, $f_{1,{\bf p}}^{\chi, o}$, Eq.~(\ref{eq:boltz_app1}) becomes
\begin{equation}
	\label{eq:boltz_app1b}
	v_z^\chi \pdv{f_{1,{\bf p}}^{\chi,o}}{z}-e E_x\pdv{f_{0,{\bf p}}^\chi}{p_x} =  -\frac{1}{\tau}  f_{1,{\bf p}}^{\chi,o}.
\end{equation}
The solution of this equation with the boundary condition of Fuchs and Sondheimer yields Eq.~(8) of the main text.

To second order in $E_x$, Eq.~(\ref{eq:boltz_app}) reads
\begin{equation}
	\label{eq:boltz_app2}
	v_z^\chi \pdv{f^\chi_{2,{\bf p}}}{z}-eE_x\pdv{f^\chi_{1,{\bf p}}}{p_x} + \frac{eE_x \Omega^\chi_{{\bf p},y}}{\hbar}\pdv{f^\chi_{1,{\bf p}}}{z}=  -\frac{1}{\tau}\left( f^\chi_{2, {\bf p}} - \frac{1}{\rho(\varepsilon_{\bf p})} \sum_{{\bf p}'}f^\chi_{2, {\bf p'}}\delta(\varepsilon_{\bf p}-\varepsilon_{\bf p'})\right).
\end{equation}
To solve this, we again assume that $\epsilon_{\bf p}^\chi$ is invariant under $p_x\to -p_x$.
Then, the second and third terms in Eq.~(\ref{eq:boltz_app2}) are even and odd functions of $p_x$, respectively (in the models we consider, $\Omega^\chi_{{\bf p},y}$ is an even function of $p_x$).
As such, if we once again decompose $f_{2,{\bf p}}^\chi$ into a part that is even in $p_x$ ($f_{2,{\bf p}}^{\chi,e}$) and a part that is odd in $p_x$ ($f_{2,{\bf p}}^{\chi,o}$), then Eq.~(\ref{eq:boltz_app2}) gives
\begin{align}
	\label{eq:boltz2_o_e}
	&v_z^\chi \pdv{f^{\chi,e}_{2,{\bf p}}}{z}-eE_x\pdv{f^\chi_{1,{\bf p}}}{p_x} =  -\frac{1}{\tau}\left( f^{\chi,e}_{2,{\bf p}} - \frac{1}{\rho(\varepsilon_{\bf p})} \sum_{\bf p'}\,f^{\chi,e}_{2,{\bf p}'}\delta(\varepsilon_{\bf p}-\varepsilon_{\bf p'})\right), \\
	&v_z^\chi \pdv{f^{\chi, o}_{2,{\bf p}}}{z} + \frac{eE_x \Omega_{{\bf p}, y}}{\hbar}\pdv{f^\chi_{1,{\bf p}}}{z} =  -\frac{f^{\chi, o}_{2,{\bf p}}}{\tau}.
\end{align}
The equation for $f_{2,{\bf p}}^{\chi,o}$ can be easily solved together with the Fuchs-Sondheimer boundary conditions; the solution is shown in Eq.~(11) of the main text.
In contrast, the equation for $f_{2,{\bf p}}^{\chi,e}$ is not amenable to an analytical solution. 
Yet, this part is not relevant to the calculation of the electric current density in Eq.~(4) of the main text.
An approximate solution is analytically obtainable using the \textit{relaxation time approximation}, i.e. setting $\sum_{\bf p'} f^{\chi,e}_{2,{\bf p}'} \delta (\epsilon_{\bf p}-\epsilon_{\bf p'})$ to zero in the collision term.  However, this approximation is hardly justified and leads to a nonconservation of the number of particles at second order, which physically unacceptable (see discussion in Sec. E). This part being not relevant in our study, we choose to not treat it in the manuscript. More discussions about it are below (see Sec. D and Sec. E).

\section{Sec. B: Derivation of the nonlinear current density in a two-dimensional Dirac metal}
\paragraph{}

In this Appendix, we calculate the part of the current density that is quadratic in the applied electric field, in a toy model for a 2D Dirac metal with time reversal symmetry and nonzero Berry curvature. 
The starting point is 
\begin{equation}
	j_{2, x}(z) = -\frac{e}{(2\pi\hbar)^2} \sum_\chi \int dp_x dp_z \, v_x^\chi f_{2,{\bf p}}^\chi,
\end{equation}
where $f_{2,{\bf p}}^\chi$ is the quadratic-in-$E_x$  part of the electronic distribution. Assuming that the energy dispersion is symmetric under $p_x\to -p_x$, $f_{2,{\bf p}}^\chi$ is given by Eq.~(11) of the main text.
Because $f_{2,{\bf p}}^\chi$ is strongly peaked at the Fermi energy,  the integral over $p_x$ and $p_z$ can be extended from $-\infty$ to $+\infty$ on each valley.
In addition, because the form of $f_{2,{\bf p}}^\chi$ depends on the sign of $v_z^\chi$,  it is mathematically convenient to change the integration variables from $\{p_x, p_z \} $ to $\{v_x, v_z\}$ using the relations
\begin{equation}
	\label{eq:pv}
	p_x =\frac{m v_x}{c\sqrt{c^2-v_x^2-(v_z-\chi\alpha_z)^2}}, \qquad p_z = \frac{m (v_z-\chi\alpha_z)}{c\sqrt{c^2-v_x^2-(v_z-\chi\alpha_z)^2}},
\end{equation}
where we hereafter omit the superscript $\chi$ in $v_x$ and $v_z$. The arguments of the square roots of Eq.~(\ref{eq:pv})  must be positive.
The Jacobian of the transformation reads
\begin{equation}
	J^\chi = 
	\begin{vmatrix}
		\pdv{p_x}{v_x} & \pdv{p_x}{v_z} \\
		\pdv{p_z}{v_x} & \pdv{p_z}{v_z} 
	\end{vmatrix}
	= \frac{m^2}{\left[v_x^2+(v_z-\chi\alpha_z)^2-c^2\right]^2}.
\end{equation}
Therefore, the nonlinear part of the current density is given by
\begin{equation}
	\label{eq:j2appB1}
	\begin{split}
		j_{2, x}(z) &= -\frac{e}{(2\pi\hbar)^2} \sum_\chi \int dv_x dv_z v_x J^\chi f_2^\chi \\
		&=  -\frac{e}{(2\pi\hbar)^2}\sum_\chi  \int dv_x\, v_x \left(\int_{v_z<0} dv_z  J^\chi f_{2}^\chi+ \int_{v_z>0}dv_z J^\chi f_2^\chi\right),
	\end{split}
\end{equation}
where we have divided the last integral in two parts because $f_{2}^\chi$ has different expressions for $v_z>0$ and $v_z<0$.


Assuming that  $|\alpha_z|\ll c$, the electronic dispersion is close to being isotropic and thus it is convenient to do the integration in Eq.~(\ref{eq:j2appB1}) in polar coordinates, 
$(v_x = v \cos{\theta}, v_z = v \sin{\theta},  dv_x dv_z = v dv d\theta$).
Then, 
\begin{align}
	\label{eq:j2app1}
	j_{2, x}(z) =  
	-\frac{e^3 E_x^2  }{\hbar (2\pi\hbar)^2} \sum_\chi \int_0^\infty dv &\left(-\int_{\pi}^{2\pi}d\theta \,\frac{\cos{\theta}}{\sin^2{\theta}} J^\chi   \frac{\partial f^\chi_0}{\partial p_x}\Omega_{{\bf p}, y}^{\chi} (a-z) e^{(a-z)/\tau v \sin{\theta}} 
	\right. \notag\\
	&\left.
	+ \int_{0}^{\pi}d\theta  \frac{\cos{\theta}}{\sin^2{\theta}} J^\chi  \pdv{f^\chi_0}{p_x}\Omega_{{\bf p}, y}^{\chi}z e^{-z/\tau v \sin{\theta}} \right),
\end{align}
where the Berry curvature in the valley $\chi$ reads 
\begin{equation}
	\label{eq:berrycurvature}
	\Omega_{{\bf p}, y}^{\chi} = \chi \frac{m \hbar^2 c^2}{2(c^2p^2 + m^2)^{3/2}} =  \chi \frac{m \hbar^2 c^2}{2(\varepsilon_{\bf p}^{\chi} - \chi \alpha_z p_z)^{3}} .
\end{equation}
Since $\partial f_0^\chi/\partial p_x$ is peaked at the Fermi energy,  we can approximate $\varepsilon_{\bf p}^{\chi} \approx \varepsilon_F$ inside the integral, and then develop the Berry curvature up to the first order in $\alpha_z$ (because $|\alpha_z p_z| \ll \varepsilon_F$ at the Fermi surface). We obtain
\begin{equation}
	\Omega_{{\bf p}, y}^{\chi} \simeq  \chi \frac{m \hbar^2 c^2}{2\varepsilon_{F}^3}\left( 1+ \chi\frac{3\alpha_z p_z}{\varepsilon_{F}}\right) =  \chi \frac{m \hbar^2 c^2}{2\varepsilon_{F}^3}+   \chi^2 \frac{3m \hbar^2 c^2}{2\varepsilon_{F}^4}\alpha_z p_z \equiv \Omega_y^{\chi}\left(1+ 3\chi \frac{\alpha_z p_z}{\varepsilon_F} \right),
\end{equation}
where $\Omega_y^{\chi} \equiv \chi \Omega_y = \chi \frac{m\hbar^2 c^2}{2 \varepsilon_F^3}$ is the Berry curvature without any tilt in the dispersion. In polar coordinates, 
\begin{equation}
	\Omega_{{\bf p}, y}^{\chi} \simeq  \Omega_y^{\chi}\left(1+ 3\chi \frac{m}{\varepsilon_F} \frac{v \sin{\theta}}{(c^2-v^2)^{1/2}}\frac{\alpha_z}{c} \right).
\end{equation}

In Eq.~\eqref{eq:j2app1}, we have written the upper bound of $v$ up as $\infty$, in apparent contradiction with the fact that the square roots in Eq.~(\ref{eq:pv}) must be positive. 
Yet, the fact that the derivative of the Fermi-Dirac distribution is peaked at the Fermi energy allows us to extend the upper bound of the $v$-integral.
Indeed, at low temperature, 
\begin{equation}
	\pdv{f_0^\chi}{p_x} = \pdv{f_0}{\varepsilon_{\bf p}^\chi}\pdv{\varepsilon_{\bf p}^\chi}{p_x} \approx -v\cos{\theta} \delta(\varepsilon_{\bf p}^\chi-\epsilon_F),
\end{equation}
where
\begin{equation}
	\delta(\varepsilon_{\bf p}^\chi-\epsilon_F) = \left| \frac{\partial \varepsilon_{\bf p}^\chi}{\partial v} \right|^{-1}_{v=v_*^\chi} \delta(v - v_*^\chi)
\end{equation}
and $v_*^\chi$ is such that $\varepsilon^\chi_{\bf p} = \epsilon_F$ when $v = v_*^\chi$. 
Expanding to first order in the tilt parameter $\alpha_z$, we get
\begin{equation}
	\label{eq:vstar}
	v_*^\chi \simeq c \sqrt{1-\frac{m^2}{\epsilon_F^2}}\left[1 + \chi\frac{\alpha_z}{c} \sqrt{1-\frac{m^2}{\epsilon_F^2}} \sin{\theta}\right]. 
\end{equation}
Insofar as the second term in the square brackets of Eq.~(\ref{eq:vstar}) is smaller than $1/2$ in absolute value, the arguments of the square roots in Eq.~(\ref{eq:pv}) are indeed positive.
In addition, we get
\begin{equation}
	\label{delta}
	\frac{\partial f_0^\chi}{\partial p_x} \simeq - \frac{v  m^2c\cos{\theta}}{\epsilon_F^3 } \left( \frac{1}{\sqrt{1-\frac{m^2}{\epsilon_F^2}}}+ 2\chi\frac{\alpha_z}{c}\sin{\theta} \right)\delta(v-v_*^\chi),
\end{equation}
where again we neglect higher order terms in $\alpha_z/c$.
The Dirac delta function makes the integration over $v$ in Eq.~(\ref{eq:j2app1}) immediate.
Afterwards, we expand the integrand up to first order in $\alpha_z$. 
The outcome is
	\begin{equation}
		\begin{split}
			j_{2, x}(z) &= \frac{e^3 E_x^2 }{\hbar (2\pi\hbar)^2} \frac{\epsilon_F}{c^2} l \sum_\chi \Omega_{y}^{\chi} \int_0^{\pi} \frac{d \theta}{\tan^2\theta}\bigg\{\left[1+ \left(\frac{z+a/2}{l}+2\sin{\theta}\right)\sqrt{1-\frac{m^2}{\epsilon_F^2}} \frac{\alpha_z \chi}{c}\right] \frac{z+a/2}{l}e^{-\frac{z+a/2}{l\sin{\theta}}} \\
			&\hspace{2cm} + \left[ 1+ \left(\frac{z-a/2}{l}-2\sin{\theta}\right)\sqrt{1-\frac{m^2}{\epsilon_F^2}} \frac{\alpha_z \chi}{c}\right]\frac{z-a/2}{l}  e^{-\frac{-z+a/2}{l\sin{\theta}}} \bigg\}\\
			&=\frac{2 e^3 E_x^2 \Omega_{ y}}{\hbar (2\pi\hbar)^2} \frac{\epsilon_F}{c^2} l \sqrt{1-\frac{m^2}{\epsilon_F^2}} \frac{\alpha_z}{c} \int_0^{\pi} \frac{d \theta}{\tan^2\theta}\bigg[\left(\frac{z+a/2}{l}+2\sin{\theta}\right) \frac{z+a/2}{l}e^{-\frac{z+a/2}{l\sin{\theta}}} \\
			&\hspace{2cm} + \left(\frac{z-a/2}{l}-2\sin{\theta}\right)\frac{z-a/2}{l}  e^{-\frac{-z+a/2}{l\sin{\theta}}} \bigg]\\
			&\equiv \frac{2 e^3 E_x^2 \Omega_{ y}}{\hbar (2\pi\hbar)^2} \frac{\epsilon_F}{c^2} l \sqrt{1-\frac{m^2}{\epsilon_F^2}} \frac{\alpha_z}{c} \,\mathcal{I}(z/l).
		\end{split}
	\end{equation}

\begin{figure}[h]
		\centering
		\includegraphics[width=0.6\textwidth]{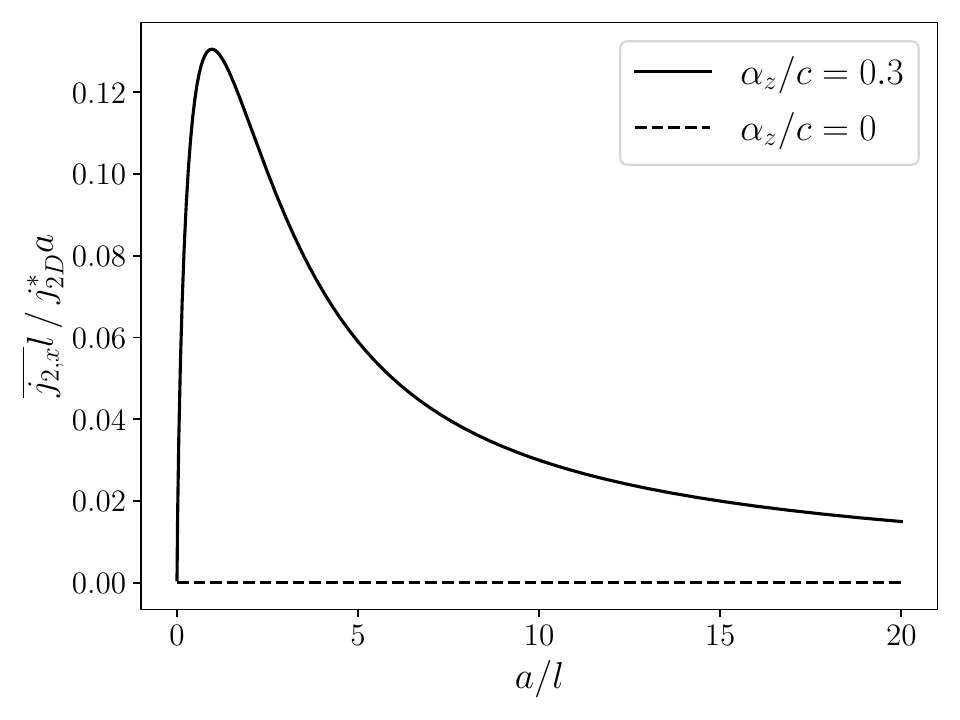}
		\caption{Average nonlinear current density as a function of the wire thickness. A tilt-independent normalization current $j_{2D}^* = j_{2D} c /\alpha_z$ is introduced, $j_{2D}^* a$ being independent of thickness. This figure shows that $\overline{j_{2,x}}\to 0$ when $a/l\to \infty$ and $a\to 0$, with a maximum that occurs when $a/l \approx 1$.}
		\label{fig:densitetot2D}
	\end{figure}

	Averaging $ j_{2, x}(z)$ over the wire thickness,
	\begin{equation}
		\overline{j_{2,x}} = \frac{1}{a}\int_{-a/2}^{a/2} dz \, j_{2,x} (z),
	\end{equation}
	using the property
	\begin{equation}
		\label{propint}
		\int_{-a/2}^{a/2} dz\, F(a/2-z) = \int_{-a/2}^{a/2} dz\, F(z+a/2)
	\end{equation}
	and following up with some algebraic manipulations, we arrive at
	\begin{equation}
		\begin{split}
			\overline{j_{2,x}}&= \frac{4 e^3 E_x^2\Omega_y}{\hbar(2\pi\hbar)^2} \frac{\varepsilon_F}{c^2}\frac{\alpha_z}{c} \sqrt{1-\frac{m^2}{\varepsilon_F^2}} \frac{l^2}{a}\int_0^{\pi} d\theta \, \frac{\cos^2{\theta}}{\sin{\theta}}\left(4\sin^2{\theta}-\left(\frac{a}{l}+2 \sin{\theta}\right)^2\text{exp}\left(-\frac{a}{l\sin{\theta}}\right)\right) \\
			&=\frac{32}{3}\frac{e^3 E_x^2\Omega_y}{\hbar(2\pi\hbar)^2} \frac{\varepsilon_F}{c^2}\frac{\alpha_z}{c} \sqrt{1-\frac{m^2}{\varepsilon_F^2}} \frac{l^2}{a} \left[1-\frac{3}{8}\int_0^{\pi}d\theta\, \frac{\cos^2{\theta}}{ \sin{\theta}}\left(\frac{a}{l}+2 \sin{\theta}\right)^2\text{exp}\left(-\frac{a}{l\sin{\theta}}\right)\right]\\
			&\equiv j_{\text{2D}} (1-I),
		\end{split}
	\end{equation}
	which is the result quoted in the main text with 
	$$j_{\text{2D}} = \frac{32}{3}\frac{e^3 E_x^2\Omega_y}{\hbar(2\pi\hbar)^2} \frac{\varepsilon_F}{c^2}\frac{\alpha_z}{c} \sqrt{1-\frac{m^2}{\varepsilon_F^2}} \frac{l^2}{a} \quad \text{and} \quad I=\frac{3}{8}\int_0^{\pi}d\theta\, \frac{\cos^2{\theta}}{ \sin{\theta}}\left(\frac{a}{l}+2 \sin{\theta}\right)^2\text{exp}\left(-\frac{a}{l\sin{\theta}}\right). $$
	The averaged current is plotted in Fig. \ref{fig:densitetot2D}.

	\section{Sec. C: Derivation of the nonlinear current density in a three-dimensional Weyl semimetal}
	\paragraph{}
	In this Appendix, we calculate the part of the current density that is quadratic in the applied electric field, in a toy model for a 3D Weyl semimetal.
	The expression for that current is
	\begin{equation}
		j_{2, x}(z) = -\frac{e}{(2\pi\hbar)^3} \sum_\chi \int dp_x dp_y dp_z \, \dot{x}_{\bf p}^\chi f_{2,{\bf p}}^\chi,
	\end{equation}
	where, under the same assumptions of the preceding Appendix,  $f_{2,{\bf p}}^\chi$ is given by Eq.~(11) of the main text.
	Because $f_{2,{\bf p}}^\chi$ is strongly peaked at the Fermi energy,  the integral over $p_i$ can be extended from $-\infty$ to $+\infty$ on each valley.
	
	Like in the previous Appendix, at first glance we are tempted to do a change of variables from $(p_x, p_y, p_z)$ to $(v_x^{\chi}, v_y^{\chi}, v_z^{\chi})$, where $v_i^{\chi} = \partial_{p_i} \varepsilon_{\bf p}^\chi$. 
	Yet, this change of variables is mathematically ill-defined, because the three velocities are not independent variables (e.g., $(v_x^{\chi})^2+(v_y^{\chi})^2+(v_z^{\chi})^2=c^2$ in the case of an isotropic dispersion).
	Instead, we perform the following transformation:
	\begin{equation}
		\begin{pmatrix}
			p_x\\p_y\\p_z
		\end{pmatrix}\to\begin{pmatrix}
			p_{\parallel}\cos\phi\\ p_\parallel\sin\phi\\ v_z^{\chi}
		\end{pmatrix}.
	\end{equation}
	The Jacobian of the transformation is 
	\begin{equation}
		\begin{split}
			\label{eq:jweyl}
			J^\chi&=p_{\parallel}\qty|\frac{\del p_z}{\del v_z^{\chi}}|
			=p_{\parallel}\qty|\frac{\del}{\del v_z^{\chi}}\qty(\frac{(v_z^{\chi}-\alpha^\chi_z)p_{\parallel}}{\sqrt{c^2-(v_z^{\chi}-\alpha^\chi_z)^2}})|
			=\frac{p_{\parallel}^2c^2}{\left[c^2-(v_z^{\chi}-\alpha^\chi_z)^2\right]^{3/2}}.
		\end{split}
	\end{equation}
	The current density is then written as 
	\begin{equation}
		\label{eq:j2app3}
		\begin{split}
			j_{2, x}&= -\frac{e}{(2\pi\hbar)^3}\sum_\chi\int_0^\infty dp_{\parallel}\int_0^{2\pi}d\phi\qty( \int_{-\infty}^0 dv_z^{\chi} \frac{v_x^{\chi} p_{\parallel}^2c^2}{\qty(c^2-(v_z^{\chi}-\alpha_z^\chi)^2)^{3/2}}f_2^\chi +\int_0^\infty dv_z^{\chi}\frac{v_x^{\chi}p_{\parallel}^2c^2}{\qty(c^2-(v_z^{\chi}-\alpha_z^\chi)^2)^{3/2}} f_2^\chi),
		\end{split}
	\end{equation}
	with the implicit constraint in the integrations over $v_z^{\chi}$ that the argument of the square root in Eq.~(\ref{eq:jweyl}) must be positive.
	
	Let us take a closer look at the integrand of Eq.~(\ref{eq:j2app3}). For $v_z^{\chi}>0$, we have
	\begin{equation}
		\begin{split}
			-\frac{v_x^{\chi} p_{\parallel}^2c^2}{\qty(c^2-(v_z^{\chi}-\alpha_z^{\chi})^2)^{3/2}}\frac{e^2E_x^2\Omega_{{\bf p}, y}^{\chi}(\del_{p_x}f_0)(z+a/2)e^{-\frac{z+a/2}{v_z^{\chi}\tau}}}{\hbar (v_z^{\chi})^2},
		\end{split}
	\end{equation}
	where 
	\begin{equation}
		\Omega_{{\bf p}, y}^{\chi}=\chi\frac{\hbar^2 p_{\parallel}\sin\phi}{2(p_{\parallel}^2+p_z^2)^{3/2}}
	\end{equation}
	is the $y$-component of the Berry curvature.
	At low temperature, 
	\begin{equation}
		\del_{p_x}f_0\simeq -v_x^{\chi}\delta(\epsilon-\epsilon_F)=-v_x^{\chi} \frac{\delta(p_\parallel-p_\parallel^*)}{\qty|\frac{\del \epsilon}{\del p_\parallel}|_{p_\parallel^*}},
	\end{equation}
	where
	$p_\parallel^*$ is the value of $p_\parallel$ such that $\varepsilon(p_\parallel^*,\phi,v_z^{\chi})=\epsilon_F$.
	We can also rewrite $v_x^{\chi}$ as a function of the new variables as 
	\begin{equation}
		v_x^{\chi}=\sqrt{c^2-(v_z^{\chi}-\alpha^\chi_z)^2}\cos\phi.
	\end{equation}
	Putting everything together and using
	\begin{equation}
		p_z=\frac{(v_z^{\chi}-\alpha_z^\chi) p_\parallel}{\sqrt{c^2-(v_z^{\chi}-\alpha_z^\chi)^2}},
	\end{equation}
	the integrand for $v_z^{\chi}>0$ becomes 
	\begin{equation}
		\label{eq:int_plus}
		-\chi \frac{e^2E_x^2\hbar}{2c (v_z^{\chi})^2}\frac{\cos^2{\phi}\sin{\phi} \left( c^2-(v_z^{\chi}-\alpha_z^{\chi})^2\right)^{3/2}\delta(p_{\parallel}-p_{\parallel}^*)}{ c^2+\alpha_y^{\chi}\sin{\phi}\sqrt{c^2-(v_z^{\chi}-\alpha_z^{\chi})^2}+ \alpha_z^{\chi}(v_z^{\chi}-\alpha_z^{\chi})} \left(z+\frac{a}{2}\right)e^{-(z+a/2)/\tau v_z^{\chi}}.
	\end{equation}
	
	The integration of Eq.~(\ref{eq:int_plus}) over $p_{\parallel}$ and $\phi$ gives
	\begin{equation}
		\label{eq:int_phi}
		-\chi \frac{e^2E_x^2\hbar}{2c (v_z^{\chi})^2} \frac{\left( c^2-(v_z^{\chi}-\alpha_z^{\chi})^2\right)^{3/2}}{c^2+\alpha_z^{\chi}(v_z^{\chi}-\alpha_z^{\chi})}\left(z+\frac{a}{2}\right)e^{-(z+a/2)/\tau v_z^{\chi}}\int_0^{2\pi}d\phi \frac{\cos^2\phi\sin\phi}{1+b\sin\phi},
	\end{equation}
	where 
	\begin{equation}
		b\equiv \frac{\alpha_y^\chi \sqrt{c^2-(v_z^{\chi}-\alpha_z^{\chi})^2}}{c^2+ \alpha_z^{\chi}(v_z^{\chi}-\alpha^\chi_z)}
	\end{equation}
	and $c^2-(v_z^{\chi}-\alpha^\chi_z)^2>0$ is required.
	The integral in Eq.~(\ref{eq:int_phi})  has three possible solutions depending on the value of $b$ ($|b|<1$, $b=1$, $|b|>1$). When the tilts are small or moderate, $|b|<1$ is the relevant regime and that is what we will focus on henceforth. Consequently,  the result of the integral is  
	\begin{equation}
		\int_0^{2\pi}d\phi \frac{\cos^2\phi\sin\phi}{1+b\sin\phi}=\frac{b^2-2+2\sqrt{1-b^2}}{b^3}\pi.
	\end{equation}
	Consequently, Eq.~(\ref{eq:int_phi}) becomes
	\begin{equation}
		-\chi \frac{e^2E_x^2\hbar}{2c (v_z^{\chi})^2} \frac{\left( c^2-(v_z^{\chi}-\alpha_z^{\chi})^2\right)^{3/2}}{c^2+\alpha_z^{\chi}(v_z^{\chi}-\alpha_z^{\chi})}\left(z+\frac{a}{2}\right)e^{-(z+a/2)/\tau v_z^{\chi}}\frac{b^2-2+2\sqrt{1-b^2}}{b^3}\pi.
	\end{equation}
	If $|\alpha_y^{\chi}|\ll c$, we may approximate
	\begin{equation}
		\frac{b^2-2+2\sqrt{1-b^2}}{b^3}\pi \approx -\frac{\pi b}{4},
	\end{equation}
	by which Eq.~(\ref{eq:int_phi}) simplifies to 
	\begin{equation}
		\chi \frac{e^2E_x^2\hbar \pi}{8c}\frac{\alpha_y^{\chi}}{(v_z^{\chi})^2} \left( \frac{ c^2-(v_z^{\chi}-\alpha_z^{\chi})^2}{c^2+\alpha_z^{\chi}(v_z^{\chi}-\alpha_z^{\chi})}\right)^2 \left(z+\frac{a}{2}\right)e^{-(z+a/2)/\tau v_z^{\chi}}.
	\end{equation}
	
	The treatment is exactly the same for $v_z^{\chi}<0$ part of Eq.~(\ref{eq:j2app3}). Putting everything together, the nonlinear current density for the valley $\chi$ reads
	\begin{equation}
		\label{eq:j2density}
		\begin{split}
			j_{2,x}^{\chi}(z) = - \chi \frac{e^3E_x^2\hbar \pi}{8c(2\pi \hbar)^3}\alpha_y^{\chi} \Bigg\{ &\int_{-c+\alpha_z^{\chi}}^0 dv_z^{\chi} \frac{1}{(v_z^{\chi})^2}\left( \frac{ c^2-(v_z^{\chi}-\alpha_z^{\chi})^2}{c^2+\alpha_z^{\chi}(v_z^{\chi}-\alpha_z^{\chi})}\right)^2 \left(z-\frac{a}{2}\right)e^{-(z-a/2)/\tau v_z^{\chi}} \\
			+&\int^{c+\alpha_z^{\chi}}_0 dv_z^{\chi} \frac{1}{(v_z^{\chi})^2}\left( \frac{ c^2-(v_z^{\chi}-\alpha_z^{\chi})^2}{c^2+\alpha_z^{\chi}(v_z^{\chi}-\alpha_z^{\chi})}\right)^2 \left(z+\frac{a}{2}\right)e^{-(z+a/2)/\tau v_z^{\chi}} \Bigg\}.
		\end{split}
	\end{equation}
	Defining the dimensionless quantities $(\Tilde{\alpha}_{y,z}^{\chi}, \Tilde{v}_z^{\chi})= (\alpha_{y,z}^{\chi}, v_z^{\chi})/c$ and $(\Tilde{z}, \Tilde{a}) = (z,a)/l$, with $l=c\tau$ as an effective bulk mean free path from impurity scattering, we have
	\begin{equation}
		\begin{split}
			j_{2,x}^{\chi}(\Tilde{z}) =  - \chi \frac{e^3E_x^2\hbar \pi l}{8c(2\pi \hbar)^3}\Tilde{\alpha}_y^{\chi} \Bigg\{& \int^{1+\Tilde{\alpha}_z^{\chi}}_0 d\Tilde{v}_z^{\chi} \frac{1}{(\Tilde{v}_z^{\chi})^2}\left( \frac{ 1-(\Tilde{v}_z^{\chi}-\Tilde{\alpha}_z^{\chi})^2}{1+\Tilde{\alpha}_z^{\chi}(\Tilde{v}_z^{\chi}-\Tilde{\alpha}_z^{\chi})}\right)^2 \left(\Tilde{z}+\frac{\Tilde{a}}{2}\right)e^{-(\Tilde{z}+\Tilde{a}/2)/ \Tilde{v}_z^{\chi}} \\
			+& \int_{-1+\Tilde{\alpha}_z^{\chi}}^0 d\Tilde{v}_z^{\chi} \left( \frac{ 1-(\Tilde{v}_z^{\chi}-\Tilde{\alpha}_z^{\chi})^2}{1+\Tilde{\alpha}_z^{\chi}(\Tilde{v}_z^{\chi}-\Tilde{\alpha}_z^{\chi})}\right)^2 \left(\Tilde{z}-\frac{\Tilde{a}}{2}\right)e^{-(\Tilde{z}-\Tilde{a}/2)/ \Tilde{v}_z^{\chi}}\Bigg\}.
		\end{split}
	\end{equation}
	If the two valleys are related by a mirror plane perpendicular to the $z$-axis ($\Tilde{\alpha}_y^+=\Tilde{\alpha}_y^-$ and $\Tilde{\alpha}_z^+=-\Tilde{\alpha}_z^-$ ), we have $j_{2,x}^+(\Tilde{z}) =j_{2,x}^-(-\Tilde{z})$.
	
	Averaging Eq.~(\ref{eq:j2density}) over the wire thickness, we get
	\begin{equation}
		\begin{split}
			\overline{j_{2,x}^{\chi}} =  - \chi \frac{e^3E_x^2\hbar \pi l}{8c(2\pi \hbar)^3}\frac{\Tilde{\alpha}_y^{\chi}}{\Tilde{a}} \Bigg\{& \int^{1+\Tilde{\alpha}_z^{\chi}}_0 d\Tilde{v}_z^{\chi} \left( \frac{ 1-(\Tilde{v}_z^{\chi}-\Tilde{\alpha}_z^{\chi})^2}{1+\Tilde{\alpha}_z^{\chi}(\Tilde{v}_z^{\chi}-\Tilde{\alpha}_z^{\chi})}\right)^2 \left[1-\left(1+\frac{\Tilde{a}}{\Tilde{v}_z^{\chi}} \right)e^{-\Tilde{a}/ \Tilde{v}_z^{\chi}}\right]\\
			-& \int_{-1+\Tilde{\alpha}_z^{\chi}}^0 d\Tilde{v}_z^{\chi} \left( \frac{ 1-(\Tilde{v}_z^{\chi}-\Tilde{\alpha}_z^{\chi})^2}{1+\Tilde{\alpha}_z^{\chi}(\Tilde{v}_z^{\chi}-\Tilde{\alpha}_z^{\chi})}\right)^2 \left[1-\left(1-\frac{\Tilde{a}}{\Tilde{v}_z^{\chi}} \right)e^{\Tilde{a}/ \Tilde{v}_z^{\chi}}\right]\Bigg\}.
		\end{split}
	\end{equation}
	When $\tilde{a}\gg 1$, the preceding expression simplifies to 
	\begin{equation}
		\label{eq:last}
		\overline{j_{2,x}^{\chi}}\simeq  \chi E_x^2 \frac{e^3 l^2}{12 a h^2 c} \frac{\alpha_y^\chi \alpha_z^\chi}{c^2},
	\end{equation}
	which is the result quoted in Eqs.~(22) and (23) of the main text. The current $j_{3D}$ introduced therein is the sum of the right hand side of Eq.~(\ref{eq:last}) over the valleys.

\begin{figure}[h]
		\centering
		\includegraphics[width=0.6\textwidth]{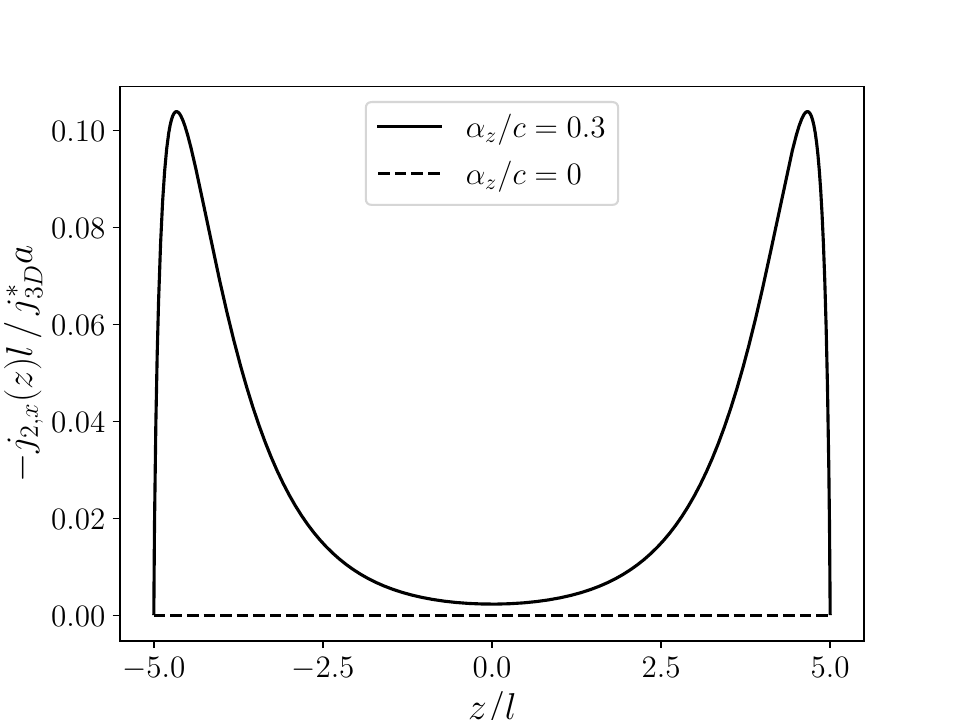}
		\caption{Spatially resolved nonlinear current density for $\Tilde{a}=10$ and $\Tilde{\alpha}_y = 0.3$. A tilt-independent normalization current $j^*_{3D} = j_{3D} / (\Tilde{\alpha}_z \tilde{\alpha}_y)$ is defined, $j_{3D}^* a$ being independent of $a$.  We see that $j_{2,x}(z)$ is even in $z$.}
		\label{fig:densitetot}
	\end{figure}

\begin{figure}[h!]
		\centering
		\includegraphics[width=0.6\textwidth]{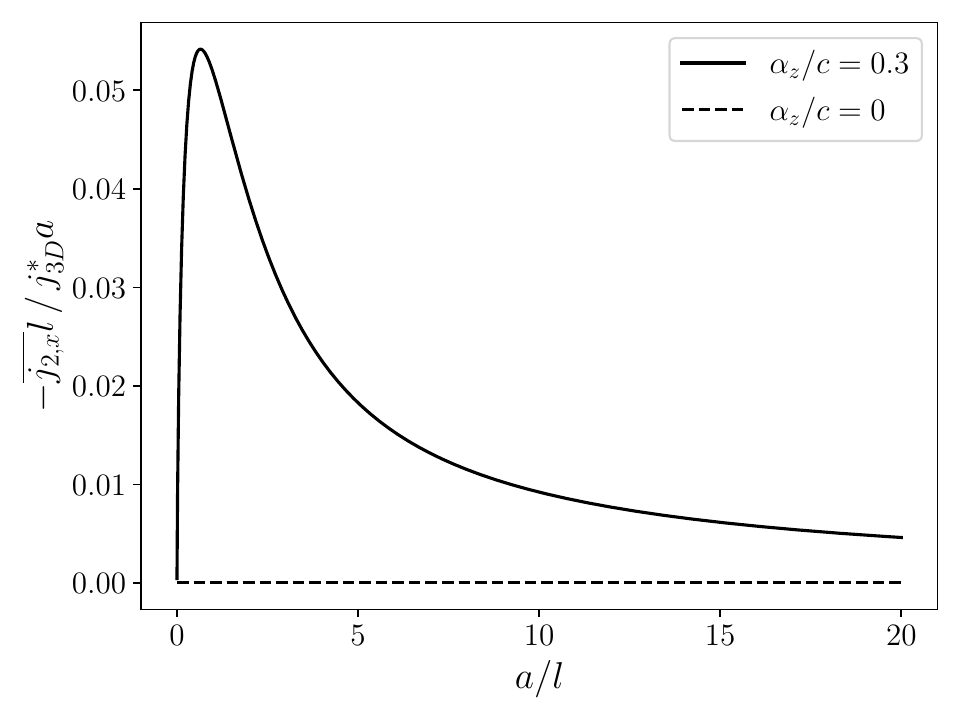}
		\caption{Average current as a function of the film thickness for $\Tilde{\alpha}_y = 0.3$. We observe  $\overline{j_{2,x}}\to 0$ when $\Tilde{a}\to \infty$ and $\tilde{a}\to 0$, with a maximum occuring for $\Tilde{a}\approx 0.5$. }
		\label{fig:couranttot}
	\end{figure}

	Figures \ref{fig:densitetot} and \ref{fig:couranttot} display $j_{2,x}(z) = \sum_{\chi}j_{2,x}^{\chi}(z)$ and $\overline{j_{2,x}} = \sum_{\chi}\overline{j_{2,x}^{\chi}}$, respectively, for the case in which the two valleys are related by a mirror plane perpendicular to the $z$-axis ($\alpha_y^\chi = \alpha_y$ independent of $\chi$ and $\alpha_z^\chi = \chi \alpha_z$). 

	\section{Sec. D: Valley polarization}
	
	
	
	In this section, we compute the valley polarization via
	\begin{equation}
		\label{eq:valdensity}
		\rho_v (z) = \sum_{\chi= \pm 1} \chi \int_{\bf p}    f_{{\bf p}}^\chi \equiv \sum_{\chi= \pm 1} \chi \, \rho^{\chi}(z),
	\end{equation}
	where $\chi$ is the valley index, $\int_{\bf p} \equiv \int d^2 p/(2\pi \hbar)^2$ and ${\bf p}$ is measured from the center of each valley.
	Also, $f_{\bf p}^\chi \simeq f_{0, {\bf p}}^\chi + f_{1, {\bf p}}^\chi + f_{2,{\bf p}}^\chi$ is the electron distribution function expanded perturbatively in powers of the electric field ($f_{n,{\bf p}}^\chi$ being proportional to $E_x^n$.).
	
	In equilibrium, $ \rho_v (z)$ vanishes because the two valleys are equally populated due to time-reversal symmetry ($\int_{\bf p}f_{0,{\bf p}}^+=\int_{\bf p}f_{0,{\bf p}}^-$).
	Likewise, there is no first-order contribution to the valley density, because $f_{1,{\bf p}}^\chi$ is an odd function of $p_x$ and thus $\int_{\bf p} f_{1,{\bf p}}^\chi = 0$.
	At second order in electric field, we write $f_{\bf p}^\chi = f_{2,{\bf p}}^{\chi, e} + f_{2,{\bf p}}^{\chi, o}$, where $f_{2,{\bf p}}^{\chi, e}$ is even under $p_x\to -p_x$ and    $f_{2,{\bf p}}^{\chi, o}$ is odd under $p_x\to -p_x$. Then, we have
	\begin{equation}
		\label{eq:rhov}
		\rho_v (z) \simeq \sum_{\chi= \pm 1} \chi \int_{\bf p}  ( f_{2,{\bf p}}^{\chi, o} + f_{2,{\bf p}}^{\chi, e})=\sum_{\chi= \pm 1} \chi \int_{\bf p}   f_{2,{\bf p}}^{\chi, e},
	\end{equation}
	because    $\int dp_x f_{2,{\bf p}}^{\chi,o}=0$.
	Thus, only the even-in-$p_x$ part of the electronic occupation, $f_{2,{\bf p}}^{\chi, e}$, can contribute to the valley density.
	This is in contrast to the electric current density displayed in Fig. 2 of the main text, to which only $f_{2,{\bf p}}^{\chi, o}$ contributes.
	
	As mentioned in Sec. A, we can obtain $f_{2,{\bf p}}^{\chi, o}$ "exactly" (in the lowest Born approximation from collisions with static impurities); see Eq.~(\ref{eq:boltz2_o_e}).
	This is so thanks to the fact that $f_{2,{\bf p}}^{\chi, o}$ is odd in $p_x$.
	Unfortunately, the same is not true for $f_{2,{\bf p}}^{\chi, e}$, because its parity in $p_x$ prevents the simplification of the collision term in Eq. (\ref{eq:boltz2_o_e}).
	Therefore, to be able to calculate $\rho_v(z)$, we need to adopt the "relaxation time approximation", by which we mean replacing the right hand side of the first line of Eq. (\ref{eq:boltz2_o_e}) with a simplified collision term
	\begin{equation}
		\label{eq:rta2}
		-\frac{f_{2,{\bf p}}^{\chi, e}}{\tau}.
	\end{equation}
	This simplification is subject to criticism, as we explain below. 
	For now, if we ignore such criticism and adopt Eq.~(\ref{eq:rta2}), we find
	\begin{equation}
		\label{eq:distribeven}
		f_{2,{\bf p}}^{\chi,e} = \left\{\begin{array}{ll} e^2E_x^2\tau^2 \pdv[2]{f_{0, {\bf p}}^\chi}{p_x}\left[1-\left(1+\frac{z+\frac{a}{2}}{\tau v_z^{\chi}}\right)e^{-(z+a/2)/\tau v_z^{\chi}}\right] - \frac{e^2E_x^2}{(v_z^\chi)^3}\pdv{f_{0,{\bf p}}^\chi}{p_x}\pdv{v_z^\chi}{p_x}\frac{\left( z+\frac{a}{2}\right)^2}{2}e^{-(z+a/2)/\tau v_z^\chi} &\text{, } v^\chi_z>0\\
			e^2E_x^2\tau^2 \pdv[2]{f_{0, {\bf p}}^\chi}{p_x}\left[1-\left(1+\frac{z-\frac{a}{2}}{\tau v_z^{\chi}}\right)e^{-(z-a/2)/\tau v_z^{\chi}}\right] - \frac{e^2E_x^2}{(v_z^\chi)^3}\pdv{f_{0,{\bf p}}^\chi}{p_x}\pdv{v_z^\chi}{p_x}\frac{\left(z-\frac{a}{2}\right)^2}{2}e^{-(z-a/2)/\tau v_z^\chi} &\text{,  } v^\chi_{z}<0.
		\end{array}\right.
	\end{equation}
	Contrary to $f_{2,{\bf p}}^{\chi,o}$, which was proportional to the bulk electronic Berry curvature,   $f_{2,{\bf p}}^{\chi,e}$ is independent of the Berry curvature.
	
	\begin{figure}[t]
		\centering
		\includegraphics[width=0.6\textwidth]{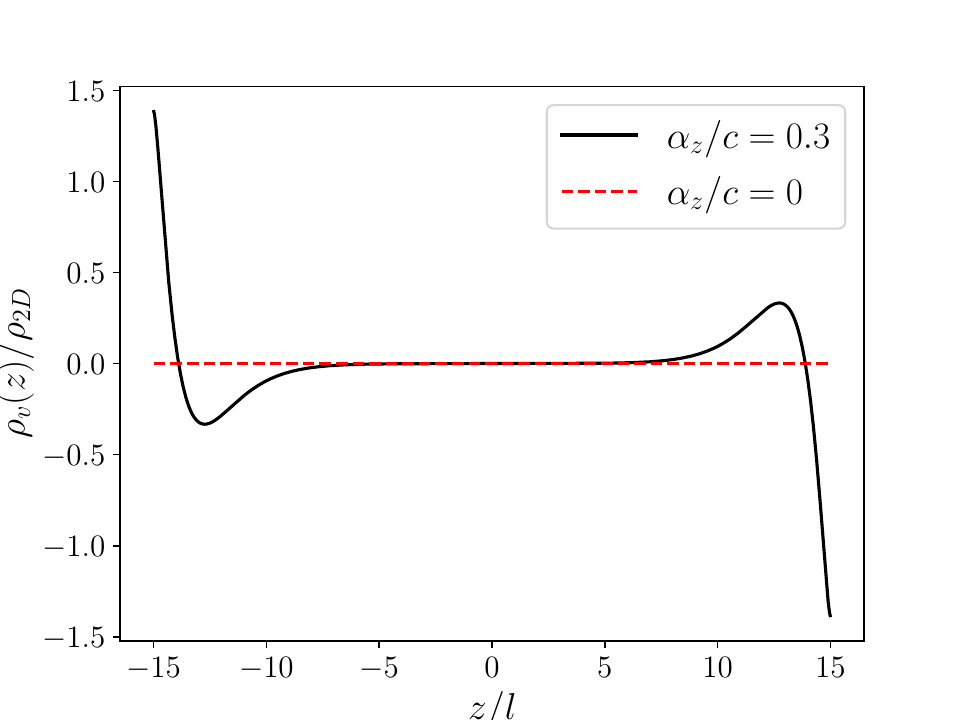}
		\caption{Valley electron density with $a/l=30$ and where $\rho_{2D}$ is defined in \eqref{eq:rho2D}. }
		\label{fig:rhovallee}
	\end{figure}

	Next, we replace Eq.~(\ref{eq:distribeven}) in Eq. (\ref{eq:rhov}). For a model of a 2D Dirac material with a gap (Eqs. (13) and (14) in the main text), we obtain
	\begin{equation}
		\label{eq:density}
		\begin{split}
			\rho^{\chi}(z)&= -\rho_{2D}\int_0^\pi  \frac{d\theta}{\tan^2\theta}\Bigg\{ \left(\frac{\left(z+\frac{a}{2} \right)^2}{l^2}e^{-\frac{(z+a/2)}{l \sin{\theta}}} + \frac{\left(z-\frac{a}{2} \right)^2}{l^2}e^{\frac{(z-a/2)}{l \sin{\theta}}} \right) \\ 
			&+ \frac{\chi \alpha_z}{c}\sqrt{1-M^2} \Bigg[\frac{\left(z+\frac{a}{2} \right)^2}{l^2}e^{-\frac{(z+a/2)}{l \sin{\theta}}} \Bigg( \frac{z+\frac{a}{2}}{l}-\frac{1}{\sin{\theta}(1-M^2)}+\frac{6M^2\sin{\theta}}{1-2M^2}   \Bigg)   \\ 
			&\hspace{2.5cm}+\frac{\left(z-\frac{a}{2} \right)^2}{l^2}e^{\frac{(z-a/2)}{l \sin{\theta}}} \Bigg( \frac{z-\frac{a}{2}}{l}+\frac{1}{\sin{\theta}(1-M^2)}-\frac{6M^2\sin{\theta}}{1-2M^2}  \Bigg) \Bigg] \Bigg\},
		\end{split}
	\end{equation}
	where $M=m/\varepsilon_F$ and 
	\begin{equation}
		\label{eq:rho2D}
		\rho_{2D} = \frac{e^2E_x^2}{2h^2c^2}\frac{l^2}{(2M^2-1)^{3/2}}.
	\end{equation}
	In the derivation of Eq. (\ref{eq:density}), we have used integration by parts to get
	\begin{equation}
		\int dp_x \, e^2E_x^2\tau^2 \pdv[2]{f_0^\chi}{p_x}\left[1-\left(1+\frac{z\pm \frac{a}{2}}{\tau v_z^{\chi}}\right)e^{-(z\pm a/2)/\tau v_z^{\chi}}\right] =  \int dp_x \,\frac{e^2E_x^2}{(v_z^\chi)^3}\pdv{f_0^\chi}{p_x}\pdv{v_z^\chi}{p_x}\left( z \pm \frac{a}{2} \right)^2e^{-(z\pm a/2)/\tau v_z^\chi}.
	\end{equation}

	Figure \ref{fig:rhovallee} displays $\rho_v(z)$ as a function of $z$.
	In the absence of a tilt in the energy dispersion, $\rho_v(z)$ vanishes at every $z$.
	In the presence of a tilt, $\rho_v(z)$ is nonzero near the edges of the wire and it is an odd function of $z$ with respect to the center of the wire. 
	The latter property can be understood from the symmetry arguments explained in the main text (indeed, $\rho^+(z) = \rho^-(-z)$ can be explained in the same way as we explained the relation $j_{2, x}^+(z) =j_{2, x}^-(-z)$ for the nonlinear current density in the main text).

	While the preceding properties of $\rho_v(z)$  are reminiscent of those of the nonlinear current density $j_{2, x}(z)$, we must emphasize important differences.
	First, $\rho_v(z)$ is unrelated to the Berry curvature, while  $j_{2, x}(z)$ scales linearly with the Berry curvature.
	Second, $\rho_v(z)$ is obtained from $f_{2,{\bf p}}^{\chi,e}$, while $j_{2, x}(z)$ follows from  $f_{2,{\bf p}}^{\chi,o}$.
	As mentioned above, we are able to obtain $f_{2,{\bf p}}^{\chi,e}$ only by making a simplifying assumption about the collision term.
	This simplification is unsatisfactory in that it leads to the violation of the number of particles at second order in the applied electric field.
	We will come back to this point below.
	Thus, the quantitative veracity of our result for $\rho_v(z)$ is debatable, although it is reasonable to believe that it is qualitatively applicable (namely, that there is a valley accumulation at the edges).
	
	In earlier work, Dyakonov \cite{dyakonov2007magnetoresistance} has shown that edge spin accumulation leads to a change in the longitudinal resistance of a conductor. 
	Let us begin by pointing out two conceptual differences between our results and those of Dyakonov.
	First, Dyakonov's theory concentrates on linear transport. In our case, the predicted result occurs only in the nonlinear transport regime.
	Second, in Ref. \cite{dyakonov2007magnetoresistance}, spin-orbit interaction is crucial in order to (i) have a spin accumulation that is linear in the electric field, and (ii) have a contribution to the electric current from the curl of the spin density.
	To realize its analogue in our system, we would need to have a valley-orbit coupling. Yet, no such coupling exists in our model. In particular, the valley degree of freedom is conserved. As a result, there is no valley accumulation in our theory at linear order in the electric field.
	Thus, an analogy between our result and that in Ref. \cite{dyakonov2007magnetoresistance} does not appear to apply.
	
	\section{Sec. E: Edge charge accumulation}
	
\begin{figure}[h]
		\centering
		\includegraphics[width=0.6\textwidth]{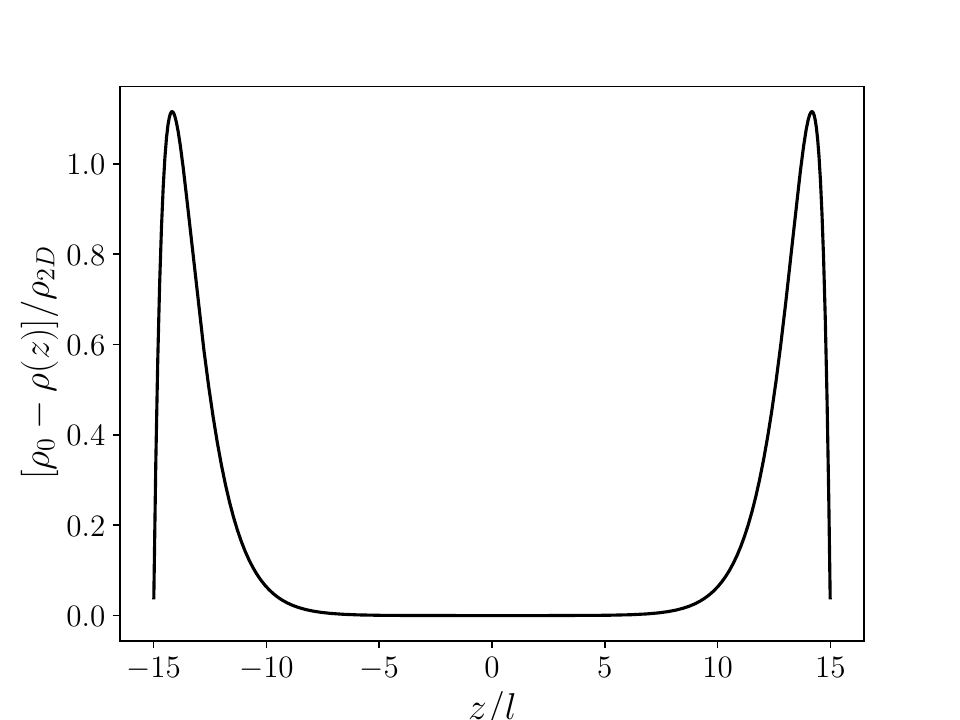}
		\caption{Spatial distribution of the nonequilibrium part of the electron density, with $a/l=30$ and $\rho_{2D}$ defined in \eqref{eq:rho2D}.}
		\label{fig:rhotot}
	\end{figure}

	If we treat $f_{2, {\bf p}}^\chi$ in the simple relaxation time approximation (Eq. (\ref{eq:distribeven})),  our theory predicts the emergence of an electric field in the direction $z$ perpendicular to the wire boundaries, near the surface of the wire.
	This can be see by computing the spatially resolved charge density, $\rho(z)$.
	Repeating the process  outlined in Sec. D, we have
	\begin{equation}
		\rho(z) = \sum_{\chi=\pm 1} \rho^\chi(z) = \rho_0-2\rho_{2D} \int_0^\pi \frac{d\theta}{\tan^2\theta} \left[\frac{\left( z+\frac{a}{2}\right)^2}{l^2}e^{-\frac{(z+a/2)}{l\sin{\theta}}}+\frac{\left( z-\frac{a}{2}\right)^2}{l^2}e^{\frac{(z-a/2)}{l\sin{\theta}}} \right],
	\end{equation}
	where $\rho_0 = \sum_\chi\int_{\bf p} f_{0,{\bf p}}^\chi$ is the equilibrium electron density (independent of $z$). 
	The out-of-equilibrium contribution to $\rho(z)$, represented graphically in Fig. \ref{fig:rhotot}, is localized near the wire boundaries and is of second order in the applied longitudinal electric field $E_x$ (recall from 
	Eq. (\ref{eq:rho2D}) above that $\rho_{2D}\propto E_x^2$).
	From Gauss' law, $\partial_z E_z = (\rho(z)-\rho_0)/\epsilon$, it follows that a transverse electric field $E_z\neq 0$ will be induced at second order in $E_x$.
	This transverse field will be localized in the vicinity of the wire boundary, where $\rho(z)\neq 0$.
	
	When combined with a Berry curvature along $y$,  $E_z\neq 0$ can lead to a Hall current in the $x$ direction. Due to time-reversal symmetry, the Hall current produced by the $E_z$ 
	field would vanish at lowest order (Eq.~(\ref{eq:anom_c}) above gives zero).
	The leading contribution is proportional to $E_z^2$, i.e. of 4th order in $E_x$. That is negligible with respect to the $E_x^2$ terms we keep in our theory.
	
	To conclude this section, we must acknowlede that Fig. \ref{fig:rhotot} evidences a fundamental flaw in the simplified calculation of $f_{2,{\bf p}}^{\chi, e}$ (based on the relaxation time approximation, Eq. (\ref{eq:rta2}) above).
	Indeed, electron number conservation is violated because $\int \rho(z) dz \neq \int dz \rho_0.$ 
	The exact collision term in the equation for $f_{2,{\bf p}}^{\chi, e}$, shown in Eq. (\ref{eq:boltz2_o_e}) satisfies charge conservation in that it vanishes upon summing over ${\bf p}$. 
	Yet, this conservation law is spuriously broken when we replace that collision term by the simpler Eq.~(\ref{eq:rta2}) above, in an attempt to obtain $f_{2,{\bf p}}^{\chi, e}$ analytically.
	Such type of defect of Eq.~(\ref{eq:rta2}) is the reason why we did not pursue the calculation of currents derived from $f_{2,{\bf p}}^{\chi, e}$ in the main text.

	
	\section{Sec. F: Comparison with earlier literature on band-geometric currents in spatially inhomogeneous systems}
	
	In this section, we compare the main finding of our manuscript to other terms of band-geometric origin that can arise in spatially inhomogeneous systems.
	
	We begin by noting that, in Ref. \cite{son2012berry}, Son and Yamamoto predicted a new contribution to the current density, which arises in the presence of both a Berry curvature and spatial inhomogeneity.
	The new term reads 
	\begin{equation}
		\label{eq:son}
		-\int_{\bf p} \epsilon_{\bf p} \boldsymbol{\Omega}_{\bf p} \times (\boldsymbol{\nabla}_{\bf r}  f_{\bf p}).
	\end{equation}
	We have not included this term in Eq. (4) of the main text. 
	
	Nevertheless, Eq. (\ref{eq:son}) does lead to a surface-localized current density along $x$ in the presence of a boundary perpendicular to $z$. In the models studied in our manuscript, Eq. (\ref{eq:son}) vanishes at zeroth and first order in the applied electric field $E_x$. 
	It vanishes at zeroth order in $E_x$ because of time-reversal symmetry.
	It also vanishes at first order in $E_x$ because, on each valley, $\epsilon_{\bf p}$ is assumed to be even in $p_x$ (this assumption allowed us to solve the Boltzmann equation without resorting to approximations in the collision term that would violate conservation laws).
	Under the same assumption, $f_{1,{\bf p}}^\chi$ is odd in $p_x$. 
	As a result, $\int_{\bf p} \epsilon_{\bf p} \Omega_y \partial_z f_{1,{\bf p}}^\chi = \partial_z \int_{\bf p} \epsilon_{\bf p} \Omega_y  f_{1,{\bf p}}^\chi =0$, where we have used the fact that $\Omega_y$ is an even function of $p_x$ in our models of interest.
	
	In contrast, there is a nonzero contribution to Eq.~(\ref{eq:son}) at second order in the electric field. It originates from $f_{2,{\bf p}}^{\chi,e}$ (the contribution from $f_{2,{\bf p}}^{\chi, o}$ vanishes for the same reason as that from $f_{1,{\bf p}}^\chi$ vanishes.)
	It is important to notice that the nonlinear current originating from Eq. (\ref{eq:son}) is qualitatively different from the one we have unveiled in the main text, because the latter originates from $f_{2,{\bf p}}^{\chi,o}$.
	
	As mentioned in preceding sections of this Supplemental Material,  we cannot reliably calculate $f_{2,{\bf p}}^{\chi,e}$.
	If we insist on using its approximate expression (Eq.~(\ref{eq:distribeven}) above) and then replace it in Eq.~(\ref{eq:son}), we do obtain a contribution to the nonlinear longitudinal current density $j_{2, x}(z)$. 
	We see that this contribution does not have the same dependence in $\tau$ as the term we have kept in the main text. This can be seen from a direct comparison between  $f_{2,{\bf p}}^{\chi,e}$ and  $f_{2,{\bf p}}^{\chi,o}$. Therefore, there is no danger of having a cancellation between the current from Eq.~(\ref{eq:son}) and the current we have discussed in the main text.
	
	On conceptual grounds, Eq.~(\ref{eq:son}) may, at least in part, be regarded as a contribution to the electric current coming from the orbital magnetic moment (OMM) of electrons.
	We base this statement on the fact that, in equilibrium (when $f_{\bf p} = f_{0,{\bf p}}$), Eq.~(\ref{eq:son}) is very closely related to the last term of Eq. (4) in Ref. \cite{xiao2006},
	\begin{equation}
		\label{eq:di}
		-\frac{1}{\beta} \int_{\bf p} \boldsymbol{\nabla}_{\bf r} \left[\log(1+e^{-\beta(\epsilon_{\bf p}-\mu)})\right] \times \boldsymbol\Omega_{\bf p} = -\int_{\bf p} f_{0,{\bf p}} \boldsymbol\Omega_{\bf p} \times (\boldsymbol{\nabla}_{\bf r}\mu),
	\end{equation}
	where we have taken $\hbar=e=1$ to match the conventions of Ref. \cite{son2012berry}.
	In Ref. \cite{xiao2006}, Eq.~(\ref{eq:di}) is the leading contribution to the electric current from the OMM of electrons. 
	Such leading contribution vanishes in systems with time-reversal symmetry. 
	When $f_{\bf p} = f_{0,{\bf p}}$, Eq.~(\ref{eq:son}) can be rewritten as
	\begin{equation}
		\label{eq:son2}
		\int_{\bf p} \epsilon_{\bf p} \frac{\partial f_{0,{\bf p}}}{\partial \epsilon_{\bf p}}\boldsymbol{\Omega}_{\bf p} \times (\boldsymbol{\nabla}_{\bf r}  \mu),
	\end{equation}
	where we have used $\boldsymbol{\nabla}_{\bf r} f_{0,{\bf p}} =-[\partial f_{0,{\bf p}}/\partial \epsilon_{\bf p}] \boldsymbol{\nabla}_{\bf r} \mu$.
	There is a suggestive resemblance between Eqs. (\ref{eq:di}) and (\ref{eq:son2}), though they are not strictly equal. 
	At first glance, one could think that Eq.~(\ref{eq:son2}) is an equivalent way to rewrite Eq. (\ref{eq:di}) as a Fermi surface (rather than Fermi sea) contribution. 
	Yet, this equivalent is not exact.
	At any rate, the resemblance between Eqs. (\ref{eq:son2}) and (\ref{eq:di}) is all the more telling, given the drastically different methods used for their derivations in Ref. \cite{son2012berry} and in  Ref. \cite{xiao2006}.
	The latter uses a rather standard semiclassical wavepacket formalism.
	The former introduces a generalized Fermi liquid theory, with "postulated" forms of commutators between operators. 
	One merit of Eq.~(\ref{eq:di}) is that it perfectly cancels the current coming from the anomalous velocity,
	\begin{equation}
		\label{eq:anom_c}
		\int_{\bf p} f_{0,{\bf p}} \boldsymbol{\Omega}_{\bf p}\times {\bf E},
	\end{equation}
	when ${\bf E} = -\boldsymbol{\nabla}_{\bf r}\mu$, i.e. when the gradient of the electrochemical potential vanishes.
	This is a physically sensible result, as there cannot be a transport Hall current in equilibrium, even if time-reversal symmetry is broken.
	In the approach of Ref. \cite{son2012berry}, Eq.~(\ref{eq:son2}) does not quite satisfy such cancellation condition. This indicates a possible problem in the results of Ref. \cite{son2012berry}.
	Conversely, one merit of the approach of Ref. \cite{son2012berry}, compared to that of Ref. \cite{xiao2006}, is that Eq. (\ref{eq:son}) appears to apply to arbitrary distribution functions, whereas only the Fermi distribution intervenes in Eq. (\ref{eq:di}).
	We have generalized Ref. \cite{xiao2006} to nonlinear responses (by inserting the nonequilibrium distribution function in the second term of Eq. (2) in Ref. \cite{xiao2006}). The extra nonequilibrium contribution to the current found that way involves $f_{2,{\bf p}}^{\chi, e}$, much like in the case of Eq. (\ref{eq:son}) above. 
	
	
	To summarize, we have dropped the contribution from Eq.~(\ref{eq:son}) for the same reason as we have dropped the contribution from the OMM of electrons in Ref. \cite{xiao2006}: both originate from $f_{2,{\bf p}}^{\chi, e}$, which cannot be calculated without resorting to approximations that violate particle number conservation.
	The current we have computed, which emerges from $f_{2,{\bf p}}^{\chi, o}$, can be reliably calculated and is qualitatively different from Eq.~(\ref{eq:son}).

\end{document}